\documentclass[10pt,journal]{IEEEtran}
\def\ps@headings{%
\def\@oddhead{\mbox{}\scriptsize\rightmark \hfil \thepage}%
\def\@evenhead{\scriptsize\thepage \hfil \leftmark\mbox{}}%
\def\@oddfoot{}%
\def\@evenfoot{}}
\makeatother
\usepackage[table,xcdraw]{xcolor}
\usepackage{booktabs} 
\usepackage{url}
\usepackage{svg}
\usepackage{etoolbox}
\usepackage{cuted}
\usepackage{lipsum}
\usepackage{amssymb}
\usepackage{array}
\usepackage{amsmath}
\usepackage{tikz}
\usepackage{dirtytalk}
\usepackage{algpseudocode}
\usepackage{graphicx}
\usepackage{caption}
\usepackage{mdframed}
\usepackage{tcolorbox}
\usepackage{comment}
\usepackage{mathtools}
\usepackage{subcaption}

\usepackage{pifont}
\usepackage{booktabs,makecell}
    
\usepackage{siunitx}

\usepackage[utf8x]{inputenc}
\usepackage{amsmath,amsthm}

\usepackage{multirow}
\usepackage[ruled,vlined]{algorithm2e}
\usepackage{fancyvrb}

\theoremstyle{definition}

\theoremstyle{plain}

\ifCLASSOPTIONcompsoc
  \usepackage[nocompress]{cite}
\else
  \usepackage{cite}
\fi
\ifCLASSINFOpdf
\else
\fi
\DeclareUnicodeCharacter{2212}{-}

\begin{document}
%
\title{\emph{B$^2$SFL}: A Bi-level Blockchained Architecture for~Secure Federated Learning-based Traffic~Prediction}

\author{Hao Guo,~\IEEEmembership{Member,~IEEE,}
        Collin Meese,~\IEEEmembership{Student Member,~IEEE,}
        Wanxin Li*,~\IEEEmembership{Member,~IEEE,} \\
        Chien-Chung Shen,~\IEEEmembership{Member,~IEEE,}
        and~Mark Nejad,~\IEEEmembership{Senior Member,~IEEE}

\IEEEcompsocitemizethanks{        
\IEEEcompsocthanksitem Manuscript received December 2, 2022; revised June 29, 2023; accepted September 20, 2023. This work is partially supported by the Fundamental Research Funds for the Central Universities under the Grant G2021KY05101, Guangdong Basic and Applied Basic Research Foundation under the Grant No. 2021A1515110286, 2021-2024, Natural Science Foundation of Shaanxi Provincial Department of Education under the Grant No. 2022JQ-639, 2022-2023, the Basic Research Programs of Taicang under Grant No. TC2022JC23, 2022-2024, XJTLU Research Development Fund under the Grant No. RDF-22-02-106, 2023-2026, and a Federal Highway Administration grant: ``Artificial Intelligence Enhanced Integrated Transportation Management System", 2020-2023. \textit{(Corresponding author: Wanxin Li.)}
\IEEEcompsocthanksitem Hao Guo is with the Research \& Development Institute of Northwestern Polytechnical University in Shenzhen, 518057, China (e-mail: haoguo@nwpu.edu.cn).
\IEEEcompsocthanksitem Wanxin Li is with the Department of Communications and Networking, Xi'an Jiaotong-Liverpool University, Suzhou, 215123, China (e-mail: wanxin.li@xjtlu.edu.cn).
\IEEEcompsocthanksitem Collin Meese and Mark Nejad are with the Department of Civil and Environmental Engineering, University of Delaware, Newark, Delaware, 19716, USA (e-mail: \{cmeese, nejad\}@udel.edu).
\IEEEcompsocthanksitem Chien-Chung Shen is with the Department of Computer and Information Sciences, University of Delaware, Newark, Delaware, 19716, USA (e-mail: cshen@udel.edu).
}
}


%
%

\markboth{IEEE Transactions on Services Computing 
,~Vol.~XX, No.~XX,~2023}%
{Shell \MakeLowercase{\textit{et al.}}: Bare Demo of IEEEtran.cls for Computer Society Journals}
\maketitle
\pagestyle{headings} 

\begin{abstract}
Federated Learning (FL) is a privacy-preserving machine learning (ML) technology that enables collaborative training and learning of a global ML model based on aggregating distributed local model updates. However, security and privacy guarantees could be compromised due to malicious participants and the centralized FL server. This paper proposed a bi-level blockchained architecture for secure federated learning-based traffic prediction. The bottom and top layer blockchain store the local model and global aggregated parameters accordingly, and the distributed homomorphic-encrypted federated averaging (DHFA) scheme addresses the secure computation problems. We propose the partial private key distribution protocol and a partially homomorphic encryption/decryption scheme to achieve the distributed privacy-preserving federated averaging model. We conduct extensive experiments to measure the running time of DHFA operations, quantify the read and write performance of the blockchain network, and elucidate the impacts of varying regional group sizes and model complexities on the resulting prediction accuracy for the online traffic flow prediction task. The results indicate that the proposed system can facilitate secure and decentralized federated learning for real-world traffic prediction tasks.

\end{abstract}
\IEEEpeerreviewmaketitle

\section{Introduction}
\label{sec:introduction}


Traffic prediction plays a significant role in improving the utilization of traffic network capacity while also helping alleviate congestion by empowering traffic management centers (TMCs) and road operators to control traffic more effectively. In addition, other applications such as route guidance and navigation systems can leverage traffic prediction methods to provide travelers with more accurate information in real time.

{Despite their strong performance in the literature, state-of-the-art traffic prediction models (e.g., deep learning (DL) models) can only perform well if trained using centralized big traffic data \cite{lee2021short}. However, centralized approaches are becoming unsuitable for emerging intelligent transportation systems (ITS) where data collection is decentralized, dynamic, and performed by heterogeneous devices (e.g., sensors, mobile phones, connected vehicles). In contrast to centralized methods, collaborative and decentralized ML approaches can better match the ITS environment, allowing models to be trained and updated online directly at the network edge to improve response time and modeling efficiency for critical ITS applications.}


{Recently, Federated learning (FL) has shown promise in facilitating privacy-preserving and collaborative machine learning across multiple application domains (e.g., healthcare \cite{xu2021federated}, internet of things (IoT) \cite{khan2021federated}, and transportation \cite{liu2020privacy}). During the FL process for traffic flow prediction, devices use their locally stored data to train location-specific traffic flow prediction models, which are periodically merged to generate a global model from the contributions of each participant. Consequently, FL can train transportation models decentrally, without necessitating data sharing, protecting the privacy of each participant's data. For these reasons, FL is a strong candidate for improving traffic prediction for emerging ITS.}


{Although the existing literature on FL-based traffic prediction is sparse, some work has experimented with FL for traffic flow prediction models \cite{zeng2021multi,liu2020privacy}. However, the existing work generally focuses on the large-scale macro FL case, where significant volumes of historical data are collected by a small number of participating organizations possessing sufficient processing power. Moreover, existing experimental setups appear to simulate FL using identical data shards across participants. This scenario is not practical given the increasing decentralization of ITS, where participants are heterogeneous with respect to data volumes and computing resources and collect location-specific data sets. The reader is referred to our recent critical review on traffic prediction methods \cite{shaygan2022traffic} for more details.}

{Most importantly, a fundamental challenge in fully decentralized federated learning is to prevent the client from reconstructing the private data of another client from its shared updates while maintaining a good level of utility for the learned models~\cite{kairouz2021advances}. Unfortunately, such local privacy approaches often have a high cost in utility and do not easily integrate with fully decentralized algorithms ~\cite{kairouz2021advances}. To address these existing drawbacks, homomorphic encryption~\cite{gentry2009fullynew} could be a potential solution. Compared to other cryptography techniques, homomorphic encryption is a form of encryption that permits users to perform computations on its encrypted data without decrypting it~\cite{gentry2009fullynew,guoehr2022}. Homomorphic encryption can be used for privacy-preserving outsourced storage and computation~\cite{guoehr2022}. It allows data to be encrypted and outsourced to a federated learning server or other computing devices. The challenge is how to integrate homomorphic encryption with federated learning procedures to preserve privacy and secure computation.}

This paper proposes the \emph{B$^2$SFL}, a bi-level blockchained architecture for secure federated learning-based traffic prediction. To address the different geographic location properties of intelligent vehicles and roadway sensors and the problems of data integrity, traceability, and system scalability, the bottom layer blockchain serves as the ledger to record the local model parameters generated from the roadside edge nodes (REN) and provide the forensics for global averaging parameters. To address the secure computation problems when the federated learning server aggregates device updates in the traditional FL procedure, we propose the DHFA (distributed homomorphic-encrypted federated averaging) algorithm and store the averaged global model parameters in the top layer blockchain to provide data security and privacy.



In summary, this paper makes the following contributions:
\begin{itemize}
    \item {We proposed a novel bi-level blockchained architecture for secure federated learning-based traffic prediction. The REN and bottom layer blockchain conduct the local model training process and store the local model parameters. The top layer blockchain stores the DHFA (Distributed Homomorphic-encrypted Federated Averaging) protected global model parameters for all regions.}
    
    \item {The new DHFA algorithm addresses the security and privacy computation problems in the parameter aggregation procedure for federated learning. In particular, we design the new partial private key distribution protocol and a partial encryption/decryption scheme to achieve end-to-end privacy-preserving features.}
    
    \item {We implemented the proposed architecture using Hyperledger Fabric, Jspaillier library, and Google Colab platform. Experimental results indicate that the system has efficient data encryption/decryption time and elucidate the impacts of regional sensor groupings on prediction accuracy for online FL models. Additionally, blockchain experiments demonstrate that transaction throughputs and latencies are suitable for real-world deployment.}
\end{itemize}

The remainder of the paper is organized as follows. We discuss related work in Section II. In Section III, we describe the system architecture. In addition, we present the problem statement, permissoned blockchain network, online federated learning of traffic prediction, DHFA protocol, and workflow of \emph{B$^2$SFL}. In Section IV, we conduct a system analysis. We discussed the security analysis and threat model in Section V. Experiments and evaluations are presented in Section VI. Section VII concludes the paper and points out future research directions.

\section{Related Work}

{We review state-of-art research work on blockchain-based federated learning schemes for IoT applications.  Chai et al.~\cite{chaihomo2021} described a hierarchical blockchain framework and a hierarchical federated learning algorithm for knowledge sharing. Vehicles learn environmental data through machine learning methods and share the learned knowledge. Jia et al.~\cite{jiabinbfl} proposed a blockchain-enabled federated learning data protection aggregation scheme with differential privacy and homomorphic encryption in an IoT environment. 
However, they do not consider the decentralized deployment of federated learning scheme.}

{Shayan et al.~\cite{shayan2020biscotti} proposed Biscotti: a fully decentralized P2P approach to multi-party ML, which utilizes blockchain and cryptographic primitives to coordinate a privacy-preserving ML process between peering clients.
Lu et al.~\cite{lubfl2020} proposed the blockchain-empowered asynchronous federated learning for secure data sharing on the Internet of Vehicles. They developed a hybrid blockchain architecture that consists of the permissioned blockchain and local Directed Acyclic Graph (DAG) to enhance the reliability and security of model parameters.
Peng et al.~\cite{pengvfchain2022} proposed a verifiable and auditable federated learning framework based on the blockchain system.
However, the blockchain system performance evaluation is missing in all proposed schemes.}

{Feng et al.~\cite{fengbafl2022} proposed BAFL: A blockchain-based asynchronous federated learning framework. The blockchain ensures that model data cannot be tampered with, while asynchronous learning speeds up global aggregation.  Li et al.~\cite{bladefl2022}  proposed a decentralized FL framework by integrating blockchain into FL. Every client broadcasts its trained model to other clients, aggregates the model with received ones, and competes to generate a block before its local training in the next round.  Qu et al.~\cite{qubfl2021} developed a decentralized paradigm for big data-driven cognitive computing (D2C), using federated learning and blockchain jointly.  However, they do not consider the internal privacy problem of federated learning.}

Qi et al.~\cite{qibfl2022} proposed blockchain-based federated learning (BFL) with a reputation mechanism for model aggregation. A reputation-constrained data contribution and reward allocation mechanism encourages data owners to participate in BFL and contribute high-quality data. Mothukuri et al.~\cite{fabricfl2021} proposed a blockchain-in-the-loop FL approach that combines classic FL and Hyperledger Fabric with a gamification component. 
Liu et al.~\cite{bfltvt2021} proposed a blockchain and federated learning model for collaborative intrusion detection in vehicular edge computing. It analyzed common attacks and shows that the proposed scheme achieves cooperative privacy-preservation for vehicles.

{Zhao et al.~\cite{zhao2020privacy} proposed the blockchain-based federated learning method that preserves privacy for IoT devices. They enforced differential privacy on the extracted features to protect customers' privacy and improve the test accuracy. Qi et al.~\cite{qi2021privacy} proposed the privacy-preserving blockchain-based federated learning for traffic flow prediction task, they also applied a differential privacy method with a noise-adding mechanism to prevent data poisoning attacks. Qi et al.~\cite{qi2021blockchain} proposed a blockchain-based, integrated homomorphic FL model to address the information silo problem. It provided gradient privacy protection by employing homomorphic encryption. Compared to our scheme, differential privacy methods need to add noise, and Qi's scheme~\cite{qi2021blockchain} still utilizes one leader node to perform the cryptography task.}


This research is the first effort to propose a bi-level blockchained architecture for secure federated learning-based traffic prediction. We designed the double-layer structure for blockchain storage and the DHFA algorithm to secure the federated averaging procedure.

\section{System Architecture}

\begin{figure}[t]
\centering
\includegraphics[width=0.49\textwidth]{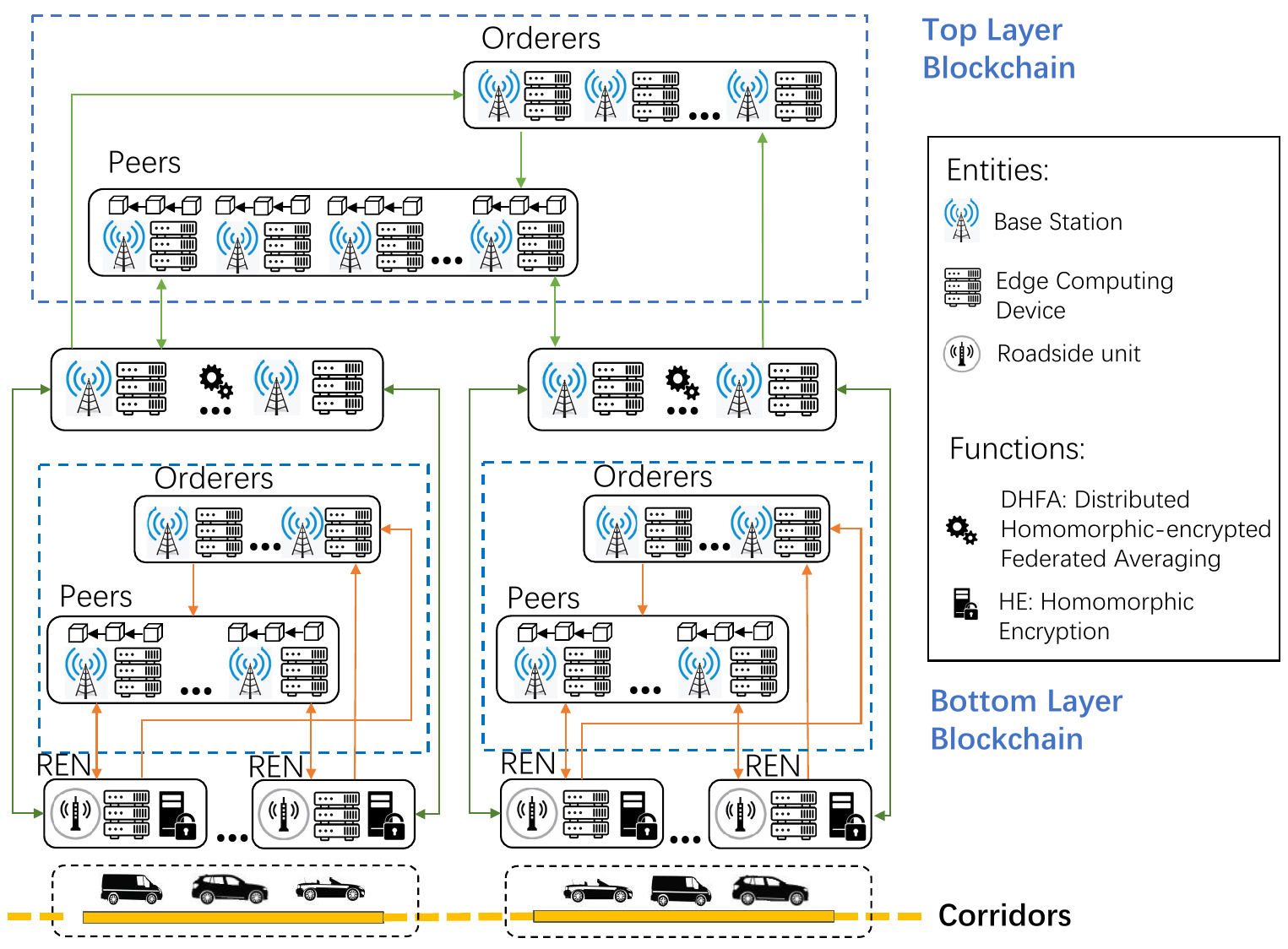}
\caption{A Bi-level Blockchained Federated Learning Architecture for Traffic Prediction.}
\label{fig:bfl}
\end{figure}

This section describes a bi-level blockchained architecture for secure federated learning-based traffic prediction with intelligent vehicles. The overall architecture is divided into four components, where the corridors layer is based on regional information, including multiple intelligent vehicles and nearby REN. The bottom layer blockchain connected to the REN serves as the ledger to record the homomorphic encrypted local model parameters. The distributed homomorphic-encrypted federated averaging scheme serves as the generator for the global parameter aggregation procedure. Finally, the top layer blockchain acts as the ledger to record the global averaging model parameters results. The system structure is shown in Fig.~\ref{fig:bfl}. We first define the following entities:





\begin{itemize}

\item Base Station: A base station is a radio receiver/transmitter that serves as the hub of the local wireless network and can be the gateway between the REN, the wireless network, and other components.

\item Edge Computing Device: Edge computing devices ($ECs$) are high-performance entities that can collaborate to accomplish the DHFA task.

\item REN: A roadside edge node (REN) collects traffic data from the static sensing area along the road. It also serves as the information source for vehicles to collect traffic information.

\item Bottom Layer Blockchain: The bottom layer blockchain serves as the ledger for regional sensors and manages the local model parameter within blockchain transactions. It forms a dynamic network based on the intelligent vehicles and REN geographic information.

\item Top Layer Blockchain: Top layer blockchain is the coordinator for the global vehicular network. It stores global federated averaging parameter and other events data.

\item {Peer Node: A peer node is a blockchain entity that conducts transaction execution, endorsement, and block validation.}

\item {Orderer Node: An orderer node is a blockchain entity that executes consensus procedures, which orders the transactions and batches them into new blocks, ensuring consistency and fault tolerance for the blockchain network.}

\end{itemize}

{In our proposed architecture, each REN, peer node, and orderer node includes a base station for communications and an edge computing device for computing and storage. Next, we define the following functions:}

\begin{itemize}

\item {Federated Learning (FL): Federated learning as a distributed machine learning approach enables nodes to collaboratively learn a shared prediction model. In our scenario, REN store and process the static sensor data to implement a federated learning process and train the local model parameter.}

\item {Homomorphic Encryption (HE): Homomorphic encryption allows users to perform computations on encrypted data without decrypting it. We utilize homomorphic encryption to protect the privacy of the training model parameter for outsourced computation and storage tasks.}

\item Distributed Homomorphic-encrypted Federated Averaging (DHFA):  DHFA is the algorithm that updates the global model parameter by calculating the average of the HE-encrypted local model parameters received from the clients. We utilize the partial homomorphic encryption scheme to protect the privacy of the model parameters. 
\end{itemize}

As shown in Fig.~\ref{fig:bfl}, the proposed framework consists of one Top Layer (TL) blockchain and multiple Bottom Layers (BLs) blockchain, in which different BL blockchains
are responsible for recording regional vehicular and local model parameters. Vehicles can communicate with each other and REN. Vehicles follow diverse roadway
routes and collect data in different corridors. Each REN operates as a worker in the federated learning process. The HE-encrypted local model parameters are stored in the bottom layer blockchain transactions. 

Additionally, the edge computing devices in the DHFA groups communicate with the REN and their edge computing devices at the corridors level. Among the $N$ edge computing devices, $N-1$ edge computing devices and one randomly chosen edge computing device execute the DHFA algorithm cooperatively. Later, the global averaging parameters are collected
by the peers and orders node, then recorded in top layer blockchain transactions. The top layer blockchain stores the global federated averaging parameters, which can be used for traffic analysis in the future.



\subsection{Problem Statement}

A typical federated learning process encompasses the federated averaging algorithm of McMahan et al.~\cite{mcmahan2017communication}, and a server (for instance, a service provider) orchestrates the training process by repeating the following steps until the training is stopped:
\begin{enumerate}
    \item Client selection: The server samples from clients meeting eligibility requirements. For instance, mobile phones can check in to the server if they are plugged in, and idle to avoid impacting the user of the device.
    
    \item Broadcast: The data users download the current model weights and the training program (e.g., a
traffic model prediction) from the server.
    
    \item {Client computation: Each selected device locally computes an update to the model by executing the federated training program, which runs the SGD (Stochastic Gradient Descent) on local data.}
    
    \item  Aggregation: The server collects an aggregate of the device updates. This stage can be the integration point for many other techniques, including secure aggregation for added privacy, and noise addition and update clipping for differential privacy.
    
    \item Model update: The server updates the shared model based on the aggregated update computed from the clients who participated in current round~\cite{kairouz2021advances}.
\end{enumerate}

A typical assumption in a federated learning system is that the participants are honest, whereas the server is honest-but-curious (HBC). As shown in Step 4 and Algorithm 1, with the client updates $(i, x_t)$ as the input, line 1 shows the system initialization procedure, and from lines 2-4, for each round of the training process, it gets the random set of $M$ clients. Line 5-7 indicates that for each client $i$ in parallel, it calculates the $\sum\limits_{k=1}^{M}\frac{1}{M}x_{1+t}^{i}$ as the $x_{t+1}$ federated averaging result.  As we can observe from Algorithm 1, the server collects an aggregate of the device updates, which may leak sensitive information. 

\textbf{Security Definition} (HBC Adversary): A federated learning system typically assumes honest participants and security against an honest-but-curious server. That is, only the server can compromise the privacy of participants' data. Therefore, no information leakage from any participants to the server is allowed \cite{10.1145/3298981}.

To address the above security issue, we propose the \textbf{secure computation problems of interest}. Secure aggregation is functionality for $n$ clients and a server. It enables each client to submit a value, such that the server learns an aggregate function of clients’ values, typically the sum value~\cite{kairouz2021advances}. We integrated the homomorphic encryption scheme into the federated averaging procedure to secure the FL training process.
 

\begin{algorithm}[t]
\label{alg:flserver}
\SetAlgoLined
\LinesNumbered
\SetKwInOut{Input}{Input}
\SetKwInOut{Output}{Output}
\Input{Client updates $(i, x_t)$}
\Output{Federated averaging $x_{t+1}$}
initialize $x_0$;\\
\For{each round $t = 1,2,...,T$}{$S_t \longleftarrow random~set~of~M~clients$;}
\For{each each client $i \in S_t$ in parallel}{$x_{t+1} \longleftarrow \sum\limits_{k=1}^{M}\frac{1}{M}x_{1+t}^{i}$}
\Comment{Federated averaging, when all clients have the same amount of data.}



\caption{The server executes the federated averaging process.}
\end{algorithm}





In this study, we propose the DHFA algorithm to conduct the global parameter averaging process based on the inputs of the local model. We improved the traditional federated
learning algorithm and enhanced the privacy-preserving concerns when performing the federated averaging procedure. Nodes on the blockchain are responsible for recording both local model parameters and global averaging parameters, and the distributed federated learning parameter aggregation procedure is realized. 

\subsection{Permissioned Blockchain Network}


{{\emph{B$^2$SFL}} adopts a permissioned blockchain network (Hyperledger Fabric) as the FL framework, which includes the entities of peer nodes and orderer nodes defined as follows~\cite{meese2022bfrt,guo2021location}. Peers nodes $P$ represent the entities that conduct transaction execution, endorsement, and block validation in both layers. Orderers nodes $O$ are the nodes that execute consensus procedures that order the transactions and batch them into new blocks, ensuring consistency and fault tolerance for the blockchain network. {In our architecture, REN and DHFA groups are the clients $C$ who interact with the bottom and top layer blockchains by sending transaction requests. In particular, we adopt Hyperledger Fabric as the platform to emulate the proposed permissioned blockchain in this paper because it offers a rich open-source community, modular design, deployment flexibility, and high parallelization capabilities due to its execute-order-validate transaction workflow \cite{10.1145/3417310.3431398}.}}


{As shown in Fig. 1, we first map each component to the permissioned blockchain architecture based on Hyperledger Fabric~\cite{li2023aggregated}. RENs, consisting of a static sensor paired with an edge computing device, collect the traffic data at a fixed position within each corridor region. Later, each REN within a regional group will participate in the local model training process and encrypt their local model parameters. Then, acting as clients, each REN will interact with peers and orderers nodes within the bottom layer blockchain. After one round of the FL process, each REN submits a transaction proposal containing the HE-encrypted local model parameter to all the edge computing
devices acting as peers for endorsements~\cite{meese2022bfrt}. Each peer evaluates the transaction proposal and sends back an endorsement to the client (REN). Each REN packages the received endorsements into a transaction and
submits it to the edge computing devices, which act as orderer nodes. The orderer nodes will execute the consensus algorithm to establish an exact order on blockchain transactions and batch them into a new block~\cite{meese2022bfrt}. {Consequently, the local model parameters of each participating edge device are stored on the respective regional BL BC.} 

{{After sufficient local models are deposited to a given regional BL BC, the DHFA group will act as the client to compute the homomorphic-encrypted federated averaging results for the global parameters and subsequently store them in the top layer blockchain.} The DHFA group will interact with the peers and the orderer nodes in the top-layer blockchain. After one round of the federated averaging process, the DHFA group will submit a transaction proposal containing the DHFA-protected global model parameter to all the edge computing devices acting as peers for endorsements. After the same evaluation process, the peer nodes will send back an endorsement to the DHFA group. The DHFA group packages all the received endorsements into a transaction and submits it to the top layer orderer nodes. The orderer nodes will establish an unambiguous order on blockchain transactions and batch them into a new block on the top layer blockchain. }

For the blockchain ledger, the data structure used for the stored model updates is as follows:
\vspace{1mm}
\begin{Verbatim}
   type ModelUpdate struct {
       FederatedID         string
       LocationID          string
       DetectorID          string
       RoundNumber         int 
       ModelParameters     HDF5
   }
\end{Verbatim}
\vspace{1mm}
The parameters within the $ModelUpdate$ data structure represent the following: $FederatedID$ refers to the name of the associated FL process (e.g., ``GRU TFP I-95"); $LocationID$ uniquely identifies the region associated with a given local model; $DetectorID$ is the identifier for a specific client ${c \in \mathcal{C}}$ within the blockchain network; $RoundNumber$ indicates the FL round index (i.e.~$i$ in $r_i$) associated with a given HDF5 file; and lastly the $ModelParameters$ field contains the HDF5 file. 

\subsection{Online Federated Learning of Traffic Prediction}

\begin{algorithm}[t]
\caption{Operations of clients $\mathcal{C}$ in region $e_i$ during round $r_i$}
\label{alg:BFRT}
\SetAlgoLined
\everypar={\nl}
\For{each client ${c \in \mathcal{C}}$ in region $e_i$ of $E$ during round $r_i$ of $R$ in parallel}
    {
        $d^{new, i}_c$ $\leftarrow$ $c$.\textsc{UPDDataSet($d^{old, i}_c$)} \Comment{Algo. \ref{alg:realtime-data-collection}}; \\
        
        \Comment{$d^{new, i}_c$ becomes $d^{old, i+1}_c$ in $r_{i+1}$}
        
        $\Vec{p_c}^{i}$ $\leftarrow$ $c$.\textsc{TrainLocalModel($\Vec{p_c}^{i-1}$, $d^{new, i}_c$)}; \\
        
        $E_{pk}(\Vec{p_c}^{i})$ $\leftarrow$ $c$.\textsc{Encrypt($\Vec{p_c}^{i}$)}; \\
        
        $tx^i_c$ $\leftarrow$ $c$.\textsc{TransactionProposal($E_{pk}(\Vec{p_i})$)}; \\
        
        $c$.\textsc{SendToPeers($tx^i_c$, $\mathcal{P}$)}; \\
        
        $etx^i_c$ $\leftarrow$ $c$.\textsc{GetEndorsedTransaction($\mathcal{P}$)}; \\
        
        $c$.\textsc{SendToOrderers($etx^i_c$, , $signature_c$, $\mathcal{O}$)}; \\
        
        $E_{pk}(\Vec{p_e}^{i})$ $\leftarrow$ $c$.\textsc{ReceiveUpdatedParams()}; \\
        
        $\Vec{p_e}^{i}$ $\leftarrow$ $c$.\textsc{Decrypt($E_{pk}(\Vec{p_e}^{i})$)}; \\
        
         $G_e^{i}$ $\leftarrow$ $c$.\textsc{UpdateModel($\Vec{p_e}^{i}$)}; \\
    }
    
    Begin round $r_{i+1}$; \\
\end{algorithm}

\begin{algorithm}[t]
\caption{ 
 Online traffic data collection and training data update of client $c$ in round $r_i$}\label{alg:realtime-data-collection}
\SetAlgoLined
\LinesNumbered
\SetKwInOut{Input}{Input}
\SetKwInOut{Output}{Output}
\Input{$d^{old, i}_c$}
\Output{$d^{new, i}_c$}
\While{within the data collection period $p$}
    {
    $d^{in, i}_c$ $\leftarrow$ $c$.\textsc{CollectData()}; \\
    $d^{new, i}_c$ $\leftarrow$ $d^{old, i}_c$ $\cup$ $d^{in, i}_c$; \\
    }
\If{$d^{new, i}_c.size > MaxDataSize$}
    {
     $RemoveSize$ $\leftarrow$ $d^{new, i}_c.size - MaxDataSize$; \\
     $d^{new, i}_c$.\textsc{RemoveOldData($RemoveSize$)}; \\
    }
\Return $d^{new, i}_c$
\end{algorithm}

\emph{B$^2$SFL} trains an online traffic flow prediction model at the network edge using data collected in real-time by traffic sensors without exchanging the collected data, resulting in dynamic and efficient-to-update models. In the proposed architecture, the clients $\mathcal{C}$ within a given geographic region $e_i$ collaboratively train a single, regional prediction model $G_e$ using Algorithm 4, leveraging the new incoming traffic data in a series of communication rounds $R=\langle r_1, r_2, \cdots, r_i, \cdots \rangle$.

{In our system, the FL process is synchronous and the RENs update the global model synchronously in each FL round.} During each round $r$, every client ${c \in \mathcal{C}}$ performs the following sequence of operations: (1) collect local traffic data; (2) train their local copy of the regional model with the collected data; (3) encrypt the local model parameters using HE-IBE and DHFA; (4) Encapsulate the local model parameters in a blockchain transaction and send it to the regional BL blockchain peers for endorsement and validation; (5) after collecting the necessary endorsements, package them into an endorsed transaction and send it to the regional BL blockchain orderers; (6) wait to receive the updated regional parameters from the DHFA group; (7) decrypt the updated regional parameters; and lastly, update their local version of the regional prediction model $G_e$. 

During initialization, each client ${c \in \mathcal{C}}$ within a given region is provided an identical model $G_e^{0}$ before the first round $r_1$. $G^0$ could optionally be a pre-trained model to jump-start the learning process. At the start of round~$r_i$, each client ${c \in \mathcal{C}}$ collects the incoming local traffic data $d^{in, i}_c$ (Algo. \ref{alg:BFRT}: line 2) for a predefined period $p$ to be combined with its local historical traffic data $d^{old, i}_c$ to form $d^{new, i}_c$ (Algo.~\ref{alg:realtime-data-collection}: lines 3-6). To mitigate overfitting of the old data and account for storage limitations, data samples in $d^{new, i}_c$ should be limited to a maximum data sample size $MaxDataSize$, where sensing data collecteddata in less recentfrom the older rounds areis excluded from the online training batchdataset in the more recentnew rounds (Algo.~\ref{alg:realtime-data-collection}: lines 6-7). {In particular, $MaxDataSize$ only controls the temporal range of the sensing measurements used for local training. On the contrary, the entire history of local and global model updates for all FL processes is stored on the respective bottom-layer blockchain to provide auditability and tamper resistance.} 

After the data collection period, all clients ${c \in \mathcal{C}}$ update their local copy of the regional prediction model $G_e$ obtained during the last FL round $G_e^{i-1}$ using $d^{new, i}_c$ (Algo. \ref{alg:BFRT}: line 4). Next, the updated local parameter vector $\Vec{p_e}^{i}$ is encrypted with the regional key through HE-IBE and is encapsulated within a transaction proposal $tx^i_c$. The proposal is sent to the peers $\mathcal{P}$ of the regional BL blockchain, who verify the transaction details and return an endorsement upon successful verification. Once the necessary endorsements are received, they are packaged into an endorsed transaction $etx^i_c$ including the client's $c$ digital signature $signature_c$ and transmitted to the orderer nodes $\mathcal{O}$. After receiving the transactions containing local model vectors $\Vec{p_e}^{i}$ from all clients ${c \in \mathcal{C}}$, the orderer nodes $\mathcal{O}$ reach agreement on their state and package them into a block for storage on the regional BL blockchain and send the new data block to the peers $\mathcal{P}$. Next, the clients ${c \in \mathcal{C}}$ wait for the DHFA group to finish the execution of Algo. 4 and subsequently receive the updated encrypted model vector $E_{pk}(\Vec{p_e}^{i})$. Lastly, the vector is decrypted, and the new parameters are used to update the local model copy of each regional client ${c \in \mathcal{C}}$, resulting in $G_e^{i}$. 
 
\subsection{Distributed Homomorphic-Enhanced Global Model Uploading Protocol}
{First, an identity-based encryption (IBE) scheme integrated with homomorphic encryption is proposed to address the centralized authority issue. The key authority center (KAC) and REN have secret coordinate information. They could use secret coordinates to calculate the slope of two points and one line to generate the edge computing device's partial private key.  It reduces the communication interactions to create the user's partial private key, reducing the risk of disclosing sensitive information in the DHFA interaction process.}
 
 
 



 \subsubsection{HE-IBE Partial Private Key Distribution Protocol}
 
 This scheme is operated by the key authority center KAC and the REN, jointly managing and distributing edge computing devices' partial private keys. KAC is responsible for $EC's$ identity authentication and authorization, distributing the certification for $ECs$ with a unique identifier, and generating the identity-based private key. KAC and REN utilized homomorphic encryption algorithms to generate the $EC's$ partial private key. The $ECs$ communicate with KAC and REN to get the partial private key. HE-IBE ensures that the KAC and REN are not known to each other so that a curious third party can not decrypt $EC's$ partial private key.
 
 Suppose KAC and REN have their confidential coordinate $(x_{KAC}, y_{KAC})$ and $(x_{REN}, y_{REN})$. We randomly pick $(r_{xKAC}, r_{yKAC})$ and then encrypts the coordinate with its public key:
 
 $c_{xKAC} = (1+n)^{xKAC}$ $r^{n}_{xKAC}$ $mod$ $n^2$;
 
 $c_{yKAC} = (1+n)^{yKAC}$ $r^{n}_{yKAC}$ $mod$ $n^2$;
 
 then send $c_{xKAC}, c_{yKAC}$ to the REN.
 
 Similarly, when REN receives the $c_{xKAC}, c_{yKAC}$, it will randomly pick the $k_{x1}, k_{y1}$, where $k_{x1} \neq k_{y1}$. REN encrypts its coordinate information:
 
  $c_{xREN} = (1+k_{x1}n)^{-xREN}$ $r^{n}_{xREN}$ $mod$ $n^2$;
 
  $c_{yREN} = (1+k_{y1}n)^{-yREN}$ $r^{n}_{yREN}$ $mod$ $n^2$;
  
  Finally, KAC and REN compute the slope according to their private coordinate information.
 
 The $EC's$ partial private key generation processes are shown as below:
 
\textit{Setup}$(1^{\lambda})$ $\longrightarrow$ $SysPara$: System initialization, it generates the system global parameter $SysPara$.
 
\textit{IKeyGen()} $\longrightarrow$ $PK_{KAC}, MK_{KAC}$: KAC generates public and private key pairs $PK_{KAC}, MK_{KAC}$.
 
\textit{RENKeyGen()} $\longrightarrow$ $PK_{REN}, MK_{REN}$: REN generates public and private key pairs $PK_{REN}, MK_{REN}$.

$KeyP_{KAC}(X_{KAC},Y_{KAC})$$\longleftrightarrow$ $KeyP_{REN}(X_{REN},Y_{REN})$: Both KAC and REN generate their coordinate $(X_{KAC},Y_{KAC})$ and $(X_{REN},Y_{REN})$, which include the public key information. Then KAC obtains the $EC's$ partial private key $SK_{ec(kac)}$, and REN interacts with KAC to get partial $SK_{ec(ren)}$ and sent it to $ECs$.

{In the last, the generated partial private keys $SK_{ec(kac)}$ and $SK_{ec(ren)}$ will be sent to edge computing devices for the DHFA scheme constructions.}
 

 
 \subsubsection{Distributed Homomorphic-encrypted Federated Averaging Scheme}
 
 As shown in Fig.~\ref{fig:secureavgmodel}, we propose a distributed homomorphic-encrypted federated averaging (DHFA) scheme which updates  global model parameter by calculating the average of HE-encrypted local model parameters received from  clients. Clients and edge computing devices $ECs$ utilize the partial homomorphic encryption scheme to obtain the global model averaging parameters. 
 
\begin{figure}[t]
\centering
\includegraphics[width=0.47\textwidth]{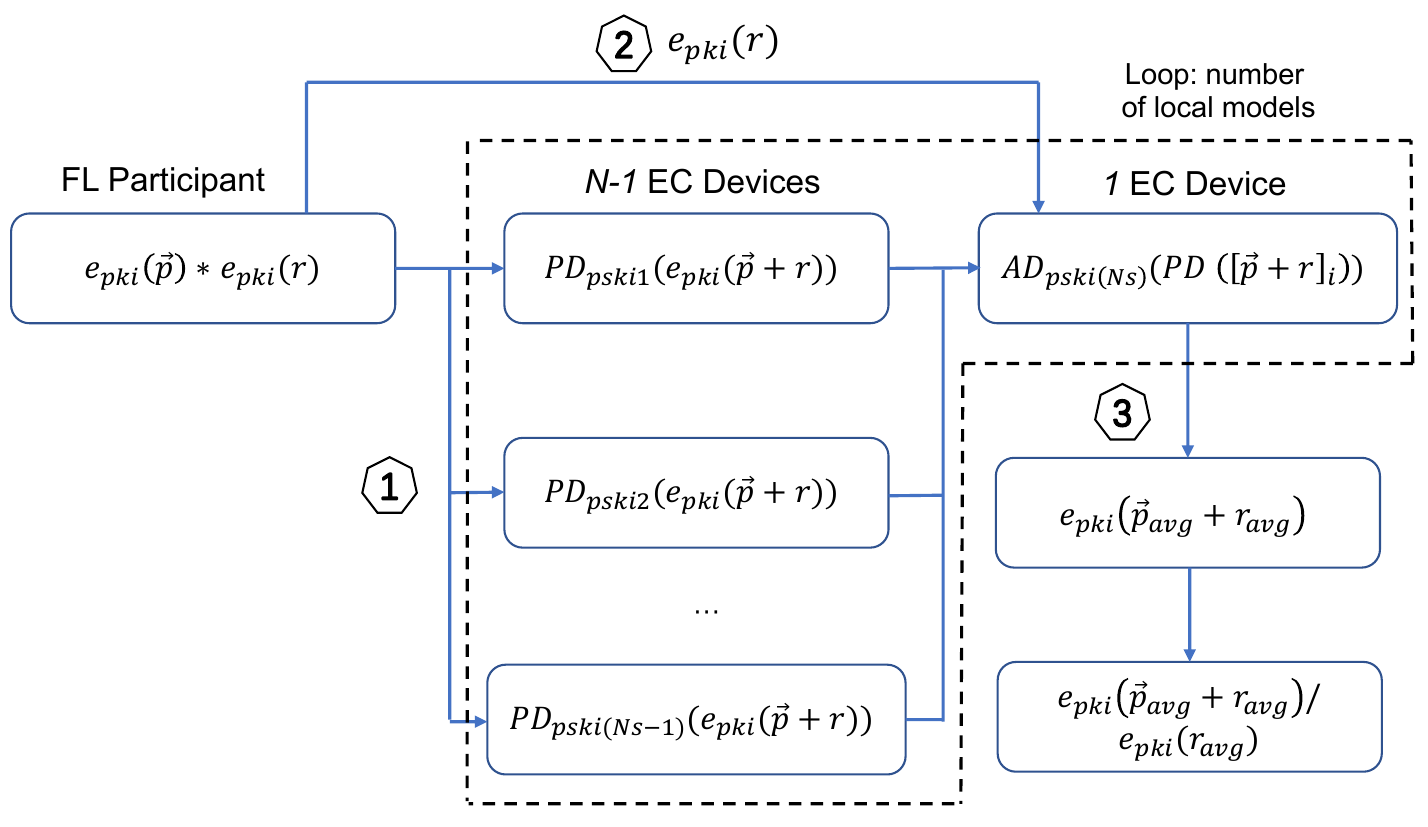}
\caption{Distributed Homomorphic-encrypted Federated Averaging Model.}
\label{fig:secureavgmodel}
\end{figure}
 
 First, the client obtains the encrypted local model parameter value $E_{pk}(\Vec{p_i})$ received from the REN, together with the encrypted random value $E_{pk}(r_i)$ utilizing homomorphic multiplication scheme, and the $r$ is the random integer value generated by the client ${c \in \mathcal{C}}$. Then the client  forwards the $E_{pk}(\Vec{p_i}) \times E_{pk}(r_i)$ to $N-1$ edge computing devices ($ECs$). The $(N−1)$ $ECs$ will perform partially homomorphic decryption operations, this process will be repeated for multiple times with all local parameters. Next, one random $EC$ executes the additive decryption operation to get summation results of the local parameter value with random number $(m + r)$. The client will send the $E_{pk}(r_i)$ value to that random $EC$.  One random $EC$ calculates the average of local parameter value with random number $(\Vec{p_{avg}} + r_{avg})$, and the client sends the $E_{pk}(r_i)$ to that random $EC$ device. 
 
 Finally, that random $EC$ removes the averaged random values $E_{pk}(r_{avg})$ from the summation average value through the reversed homomorphic addition operation, and gets encrypted average local value $E_{pk}(\Vec{p_{avg}})$ since the random integer value is generated by the client. Lastly, the $ECs$ send the encrypted  $(\Vec{p_{avg}})$  to clients. The clients could decrypt the $(\Vec{p_{avg}})$ using the symmetric HE keys and conduct the operation for the next round of the federated training process.

 \begin{algorithm}[t]
\label{alg:secureavg}
\SetAlgoLined
\LinesNumbered
\SetKwInOut{Input}{Input}
\SetKwInOut{Output}{Output}
\Input{Local model vectors $[ \Vec{p_l}]_1, [ \Vec{p_l}]_2,...,[ \Vec{p_l}]_{N_c}$}
\Output{Aggregated Global model vectors for clients $[ \Vec{p_g}]_1, [ \Vec{p_g}]_2,...,[ \Vec{p_g}]_{N_c}$}
Client generates $N_c$ random values  $R_1, R_2,...,R_{N_c}$ and encrypted them with public keys;\\
\For{$i \leq N_c$}{Client gets $[\Vec{HA_i}]_i \longleftarrow [\Vec{p_l}]_i \cdot [R_i]_i$; \\ 
}
Client sends $[[\Vec{p_l}]_1 \cdot [R_1]_1, [\Vec{p_l}]_2 \cdot [R_2]_2,..., [\Vec{p_l}]_{N_c} \cdot [R_{N_c}]_{N_c}$ to the $N-1$ ECs;\\
\For{$i \leq M$}{$N-1$ ECs partially decrypts the $PD[\Vec{HA_i}]_{i}$ using $sk_{ec}$ and obtains $(\Vec{p_l}_1 + R_1),...,\Vec{p_l}_{N_c} + R_{N_c})$;\\One random EC generates the $\Vec{p}_{sum_i} = \frac{\sum\limits_{k}{\Vec{p_l}}_k + \sum\limits_{k}{R_k}}{N_c}$;\\}

\For{$i \leq N_c$}{ECs encrypts $\Vec{p}_{sum_i}$ utilizing the public key $pk_i$;}
One random EC gets the encrypted value $[[\Vec{p}_{sum}]_1,[\Vec{p}_{sum}]_2,...,[\Vec{p}_{sum}]_{N_c}]$;\\
Client sends the $E_{pk}(r_i)$ to one random EC device;\\
\For{$i \leq N_c$}{EC gets $[\Vec{p_g}]_i \longleftarrow [\Vec{p}_{sum}]_i \div [\frac{\sum\limits_{k}^{m} (R_k)}{N_c}]_i$;}



\caption{Distributed Homomorphic-encrypted Federated Averaging Scheme}
\end{algorithm}
 
 Algorithm 4 described the distributed secure averaging global parameter generation process. The local model vectors $[\Vec{p_l}]_{N_c}$ are the inputs and the aggregated global model vectors for clients $[\Vec{p_g}]_{N_c}$ are outputs. The $ECs$ have the partial private keys $sk_{ec}$. The detailed operation of Algorithm 4 is as follows:
 \begin{enumerate}
     \item  Client receives the HE-encrypted local model vectors from the REN. Local model vector encrypted with HE public key of $i-th$ client is described as $[\Vec{p_l}]_{i}$. Next, Client generates $N_c$ random values $R_{N_c}$ together with local model vector and encrypts them utilizing  public key, shown as line 1;
      \item The client conducts the  homomorphic multiplication operations with HE-encrypted local model parameters $[\Vec{p_l}]_{i}$ together with the encrypted random vectors $[R_i]_{i}$ in line 3 when $i \leq N_c$. The homomorphic multiplication result is represented as $[\Vec{HA_i}]_{i}$
This process is repeated for $N_c$ local model parameters. Client sends $[\Vec{HA_{N_c}}]_{N_c}$ to the $N-1$ $ECs$, as showing in line 5;

\item The $N-1$ $ECs$ partially decrypts the $PD[\Vec{HA_i}]_{i}$ utilizing the partial private key $sk_{ec}$ and gets ${\Vec{p_l}}_{N_c} + R_{N_c}$.
This process will repeat for $M$ times when all the local parameters have been processed.

\item Next, one randomly chosen $EC$ will generate all the encrypted summation values and divides the sum by total number of parameters $N_c$ to get the vector including the average parameter represented as $\Vec{p}_{sum_i}$ in line 8;

\item One random $EC$ encrypts $\Vec{p}_{sum_i}$ using the homomorphic public keys $pk_i$ in lines 9 and then gets the encrypted values $[[\Vec{p}_{sum}]_1,[\Vec{p}_{sum}]_2,...,[\Vec{p}_{sum}]_{N_c}]$ in line 13;
\item The client sends the $E_{pk}(r_i)$ to one random $EC$ device in line 14; 
\item In last step, the random $EC$ gets the DHFA-encrypted global model parameters by
conducting the reversed homomorphic multiplication operation to remove the random parameter $[\frac{\sum\limits_{k}^{m} (R_k)}{N_c}]_i$ as shown in line 16.
 \end{enumerate}
 
 After the client and $ECs$ finish the global parameter averaging procedure by performing the proposed DHFA algorithm, the one random $EC$ sends the updated global model parameters to the peer nodes in the top layer blockchain. When REN start the next round of the federated learning process and send newly trained local model
parameter to clients. Clients and $ECs$ will execute the DHFA algorithm to generate the global averaging parameters again, and $ECs$ will send back the encrypted averaging parameter   $(\Vec{p_{avg}})$  to clients, and the clients can decrypt the $(\Vec{p_{avg}})$ using the symmetric HE keys.


\subsection{Workflow of B$^2$SFL}
\begin{figure}[t]
\centering
\includegraphics[width=0.49\textwidth]{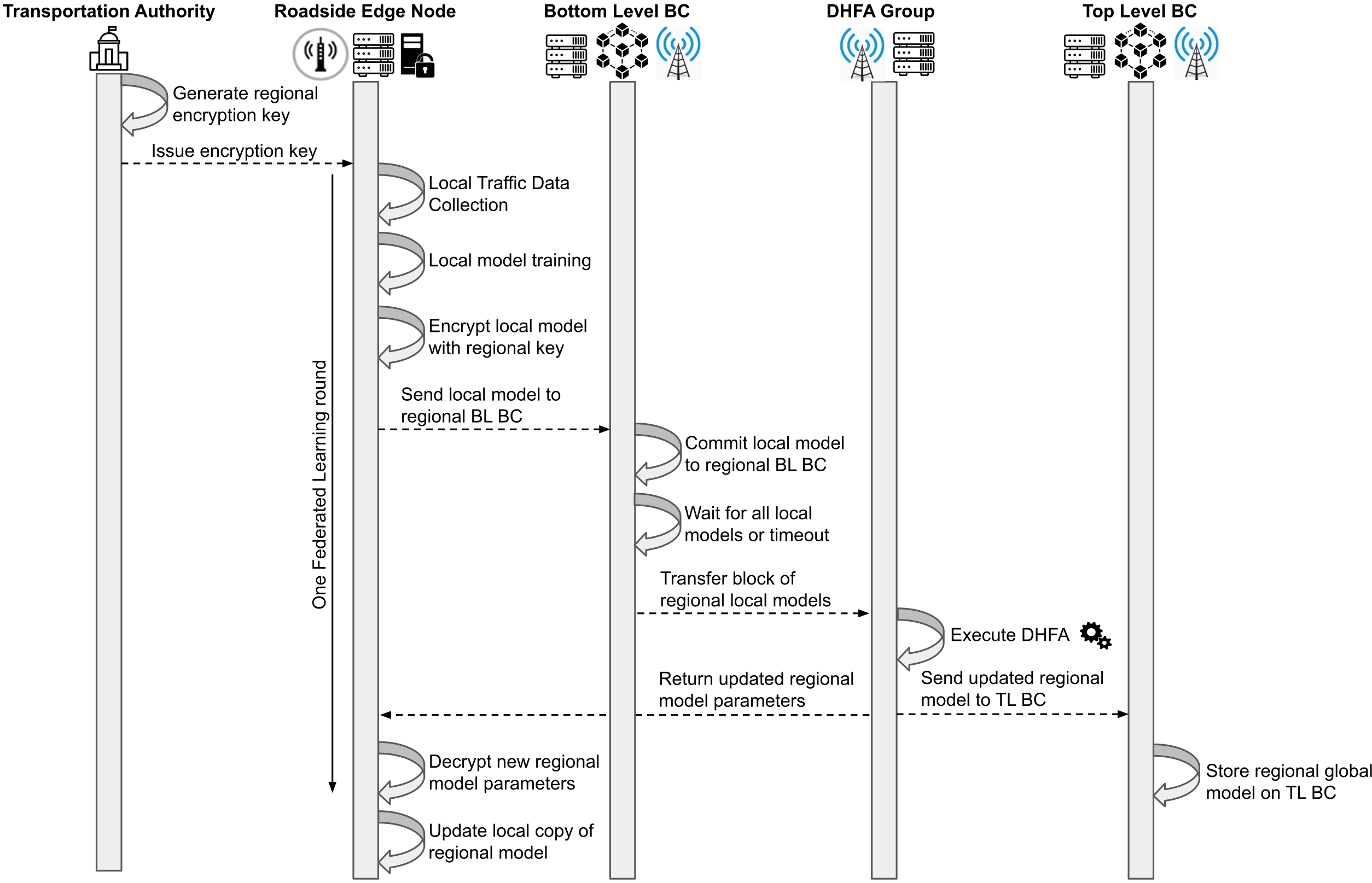}
\caption{{Workflow of a bi-level blockchained architecture for secure federated learning-based traffic prediction.}}
\label{fig:flwf}
\end{figure}

{Figure \ref{fig:flwf} illustrates the federated learning workflow for online and location-aware traffic flow prediction within the proposed architecture. First, during initialization, the public/private key is generated for each region using the proposed HE-IBE (holomorphic encryption - identity-based encryption) and is distributed to the participants. Then, throughout the FL process, the RENs within a pre-defined geographic region continuously collect local traffic data using their sensors (e.g., loop detector, LIDAR). During an FL round, after collecting sufficient local traffic data, each REN will leverage a combination of newly collected and historical data to train its local version of the global FL model. After training, the new model is encrypted using the regional key to protect the privacy and security of the local model parameters.}

Next, REN encapsulates encrypted local model parameters within a blockchain transaction, which is sent to the peer nodes of their regional BL blockchain. The peer nodes validate the transaction and return a signed endorsement to REN. After receiving the necessary endorsements, the REN packages them into a digitally signed transaction and sends the transaction to the orderer nodes. After that, the orderer nodes will enact the consensus process, package newly submitted local models into a new block, and transfer the new block to the peers in their respective BL BC, where it is used to update the ledger state.

{After block generation, the new block of updated regional global models is retrieved from the BL blockchain by the DHFA nodes acting as clients. To ensure fault tolerance of the synchronized FL process, we propose a timeout system parameter to account for communication or system outages for RENs. Specifically, the timeout value can be implemented at the DHFA node level using a client application, or implemented within the business logic of a respective BL blockchain. Once the local models of all participating RENs have been committed to the regional BL BC that manages the FL process, or when the current round timeout value is reached, the DHFA nodes will retrieve the local models and execute FedAVG.}

{After retrieving the local models, the proposed HE-FedAVG (Alg. 4) is executed to average the encrypted local model parameters into an updated version of the region-specific model. The newly encrypted regional model parameters are then transferred to the associated RENs. At the same time, the new parameters are also encapsulated in a blockchain transaction and are subsequently stored on the TL blockchain using the same transaction processing flow. Once the REN receives the updated model parameters, they are decrypted and used to update the REN's local copy of the regional prediction model, and the process repeats. }







\section{System Analysis}

We introduce and prove two propositions in this section to support the distributed homomorphic-enhanced global model uploading protocol to utilize the secret coordinates calculating the slope, and generating partial private keys without leaking any sensitive information.

\textbf{Proposition 1.} {\it{The secure slope symbol calculation process can preserve the information leakage during the partial private key generation procedure.}}
\begin{proof}

We define the
\begin{align}
P(X)=
\left\{\begin{matrix}
\textrm{+1}, X>1 & \textrm{positive slope} \\ 
\textrm{-1}, X<1 & \textrm{negative slope}\\ 
\end{matrix}\right.
\label{eq:platoonStrategy}
\end{align}

If $\frac{x}{y} > 1$ ($x$ and $y$ are point coordinates), then we have $\frac{x}{y} > \frac{x+m}{y+m} > 1$, else if $\frac{x}{y} < 1$, then we have $\frac{x}{y} < \frac{x+m}{y+m} < 1$. As a result, we get the following results $(m > 0)$:
\begin{align}
    P (\frac{x_a}{x_b}) =  P (\frac{x_a + m}{x_b + m}),
\end{align}
\begin{align}
    P (\frac{y_a}{y_b}) =  P (\frac{y_a + m}{y_b + m}).
\end{align}
If we set:
\begin{align}
P(X' - Y')=
\left\{\begin{matrix}
\textrm{+1}, X'>Y' & \textrm{positive case} \\ 
\textrm{-1}, X'<Y' & \textrm{negative case}\\
\end{matrix}\right.
\label{eq:platoonStrategy}
\end{align}
Then we can indicate that:
\begin{align}
    P (\frac{x_a}{x_b}) =  P' (x_a - x_b),
\end{align}
\begin{align}
    P (\frac{x_b}{x_a}) =  P' (x_b - x_a),
\end{align}
and
\begin{align}
    P (\frac{y_a}{y_b}) =  P' (y_a - y_b),
\end{align}
\begin{align}
    P (\frac{y_b}{y_a}) =  P' (y_b - y_a).
\end{align}
As a result, partial private key distribution protocol can calculate the symbol of the slope correctly, and there is no information leakage during partial private key generation procedure.
\end{proof}

\textbf{Proposition 2.} {\it{The protocol of two private points secret equation of a line can correctly solve the equation of a line through two private points without leaking sensitive information.}}
\begin{proof}

\begin{align*}
    Enc(A, (x_a, y_a)) = ((1 + n)^{x_a} r^{n}_{x_a} mod n^2, \\(1 + n)^{y_a} r^{n}_{y_a} mod n^2)
\end{align*}

\begin{align*}
    Enc(B, (x_b, y_b)) = ((1 + k_{x1}n)^{x_b} r^{n}_{x_b} mod n^2, \\(1 + k_{y1}n)^{y_b} r^{n}_{y_b} mod n^2)
\end{align*}

Here $k_{x1}$ and $k_{y1}$ are two different random numbers.

\begin{align*}
    Enc(A, (x^{k_{x1}}_a, y^{k_{y1}}_a)) = ((1 + n)^{k_{x1} x_a} r^{k_{x1}n}_{x_a} mod n^2, \\(1 + n)^{k_{y1} y_a} r^{k_{y1}n}_{y_a} mod n^2)
\end{align*}

Next, $B$ calculates the $Eny(k_{x1}\Delta x)$ and $Eny(k_{y1}\Delta y)$ through the $(c_{x_a}, c_{y_a})$:

we have
\begin{align*}
    k_{x1} \Delta x =  k_{x1} (x_a - x_b)
\end{align*}
\begin{align*}
    k_{y1} \Delta y =  k_{y1} (y_a - y_b)
\end{align*}

\begin{align*}
    c_{k_{x1}\Delta x} = (1 + k_{x1}n)^{x_a - x_b} ((r_{xb})^{n-x_b} r_{x_a}^{k_{x1}})^n mod\ n^2
\end{align*}

\begin{align*}
    c_{k_{y1}\Delta y} = (1 + k_{y1}n)^{y_a - y_b} ((r_{yb})^{n-y_b} r_{y_a}^{k_{y1}})^n mod\ n^2
\end{align*}

After that, we get the
$(c_{y1}, c_{x1}, (c_{y2}, c_{x2},..., (c_{yl}, c_{l})$;

and then $A$ can compute the slope $K$:
\begin{align*}
  \frac{L(c_{y1}^{\lambda} mod n^2)}{L(c_{x1}^{\lambda} mod n^2)}   \times \frac{L(c_{y1}^{\lambda} mod n^2)}{L(c_{x1}^{\lambda} mod n^2)} \times \cdot \cdot \cdot \frac{L(c_{y1}^{\lambda} mod n^2)}{L(c_{x1}^{\lambda} mod n^2)} =\\
  \frac{L(c_{y1}^{\lambda} mod n^2) \cdot L(c_{y1}^{\lambda} mod n^2) \cdot L(c_{y1}^{\lambda} mod n^2) }{L(c_{x1}^{\lambda} mod n^2)\cdot L(c_{y1}^{\lambda} mod n^2) \cdot L(c_{y1}^{\lambda} mod n^2)} = \\
  \frac{\Delta y}{\Delta x} (k_{x1} \cdot k_{x2} \cdot ... \cdot k_{xl} = k_{y1} \cdot k_{y2} \cdot ... \cdot k_{yl} ) =\\ K
\end{align*}
 
 $K$ is the slope of the line which goes through $A$ and $B$, and $A$ computes the line equation:
 
 \begin{align*}
     y = K(x - x_a) + y_a
 \end{align*}
 
 Similarly, when $B$ receives the $K$, it also computes the line equation:
 
\begin{align*}
     y = K(x - x_b) + y_b
 \end{align*}
 
The information $A$ sends to $B$ is ciphertext during the partial private key construction process, and $B$ cannot decrypt the information since it does not have the private key. Also, $A$ cannot calculate the coordinate information of $B$. The secret information of either $A$ or $B$ was not leaked during the partial private key generation process, the slope of the linear equation was correctly solved, and the system correctness was proved.
\end{proof}

\section{{Security Analysis}}

In traditional FL,  failure of a participant or central server can negatively impact FL performance. In this section, we discuss threat model from two perspectives: server vulnerabilities and participant vulnerabilities.
\subsection{Single Point of Failure Attack}
The central server is a vital coordinator for FL, collecting and aggregating all local model updates. Suppose the central server suffers a typical single point-of-failure attack. In that case, the execution of the model update aggregation will fail, and the new global model will not be assigned to the participant for subsequent local model training, which means the entire FL algorithm terminates. Our proposed scheme can avoid this issue since the FL process is done by the DHFA, which has a distributed architecture. Compared to one centralized FL server, $(N−1)$ $ECs$ will perform partially homomorphic decryption operations, this process will be repeated for multiple times with all local parameters.
\subsection{Member Inference Attack}
Since the central server knows local model updates from all participants, the central server can record the parameter of the local model. The central server can further utilize the parameter information of the local model to perform membership inference attacks to steal the local data sets of the participants. In this case, participants have no way of knowing whether their parameter information is being logged by the central server. Our proposed system can solve this problem since the parameter, and local models are encrypted with HE, which can not be directly exposed. HE allows data to be encrypted while performing computation on other computing devices.
\subsection{Label Flipping Attack} 
During the execution of the FL process, malicious participants may modify the tags of the local data set to provide low-quality model updates, which will affect the entire federated learning phase. For instance, a vehicle can change traffic flow data at time $t$, affecting traffic flow prediction at that time $t+1$. However, such an attack is not possible in our system. All the local parameter and data is stored in the bottom blockchain as the permanent record. Also, once the encrypted model has been uploaded to the top-layer blockchain, which can not be modified again to change the label. Blockchain is immune to the label flipping attack.}





\section{Experiments and Evaluations}

\subsection{Performance of Blockchain Network}
\subsubsection{Setup} The blockchain modular was developed with Hyperledger Fabric blockchain v2.2. It was deployed and experimented on the virtual machine with 3.70 GHz Intel i9-10900K processor and 24GB of memory. We instantiated four peers and five orderers using the Raft consensus algorithm. In the experiments, we used Hyperledger Caliper tool to measure the blockchain performance in two metrics: transaction throughput and transaction latency~\cite{guo2022hierarchical}.

Transaction throughput quantifies the number of transactions per second (tps) which can be successfully processed by the blockchain network, while the latency indicates the average running time of transactions from initial construction by the clients until successfully committed to the ledger. In all the experiments, we analyze both metrics by increasing transaction send rates which indicate the number of input transactions per second by the blockchain clients. Notably, the number of clients for each bottom layer blockchain is usually less than seven due to the bi-level design.


\subsubsection{Transaction Throughput}

\begin{figure}[t]
\centering
\includegraphics[width=0.24\textwidth]{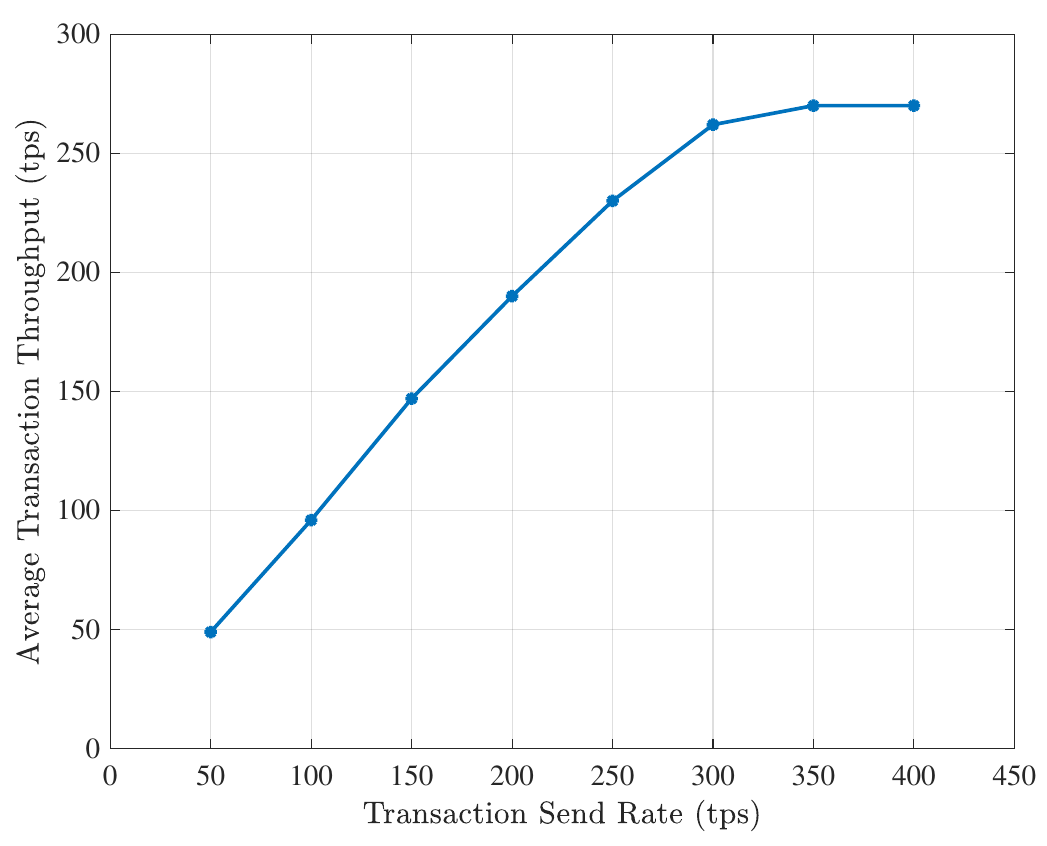}
\caption{Transaction throughput vs. transaction send rate.}
\label{fig:throughput}
\end{figure}

In Fig. \ref{fig:throughput}, we illustrate the throughput of our blockchain network. We report the average transaction throughput over multiple testing cycles to quantify performance. It shows that the transaction throughput increases as the send rate increases to 400 tps. However, the performance levels off at send rates above 350 tps, and throughput remain relatively constant at 270 tps. This indicates that the maximum network throughput is about 270 tps.

\begin{figure}[t]
\centering
\includegraphics[width=0.24\textwidth]{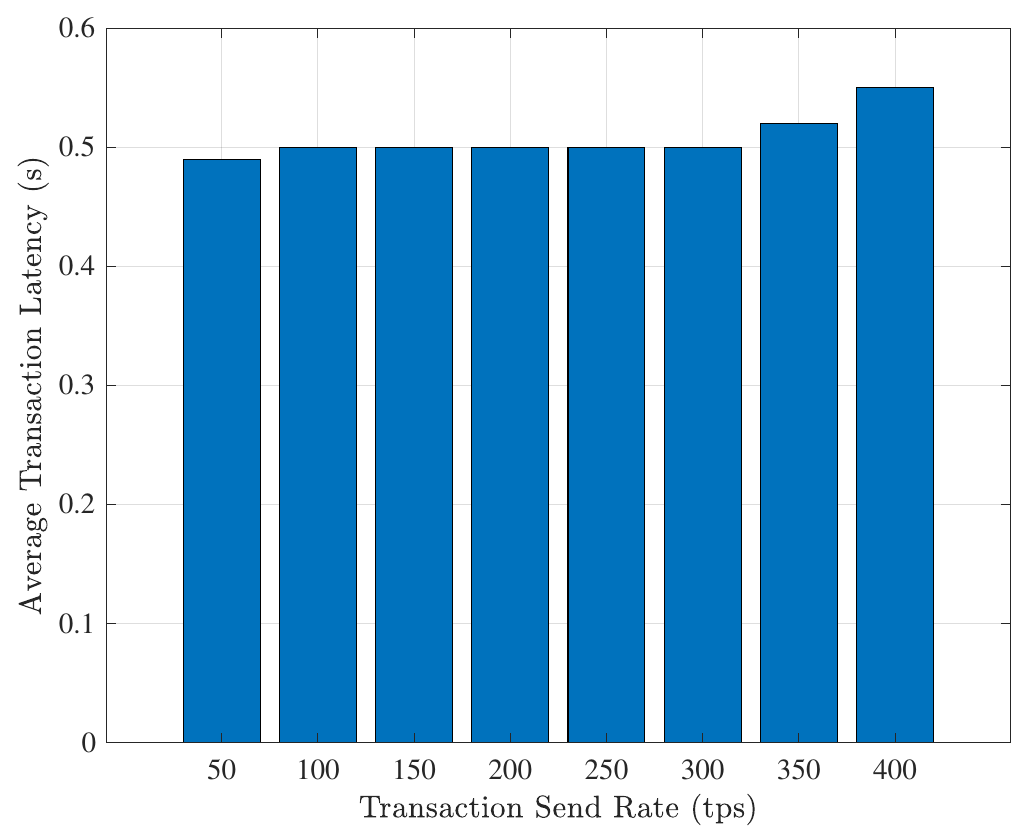}
\caption{Transaction latency vs. transaction send rate.}
\label{fig:latency}
\end{figure}

\subsubsection{Transaction latency}

{In Fig. \ref{fig:latency}, we illustrate the latency of our blockchain network. The average transaction latency remains  at 0.5 s when the transaction sends rate is below 300 tps. After passing 300 tps transaction send rates, the transaction latency increases slowly. The results indicate our blockchain network can handle the transaction requests from a certain number of RENs in a corridor. Compared to the centralized solution, our scheme is feasible in transaction latency for traffic management systems.} 






\subsection{Homomorphic Encryption Execution Comparison Result}
\label{section:HE cost}

\begin{figure}[t]
\centering
\includegraphics[width=0.24\textwidth]{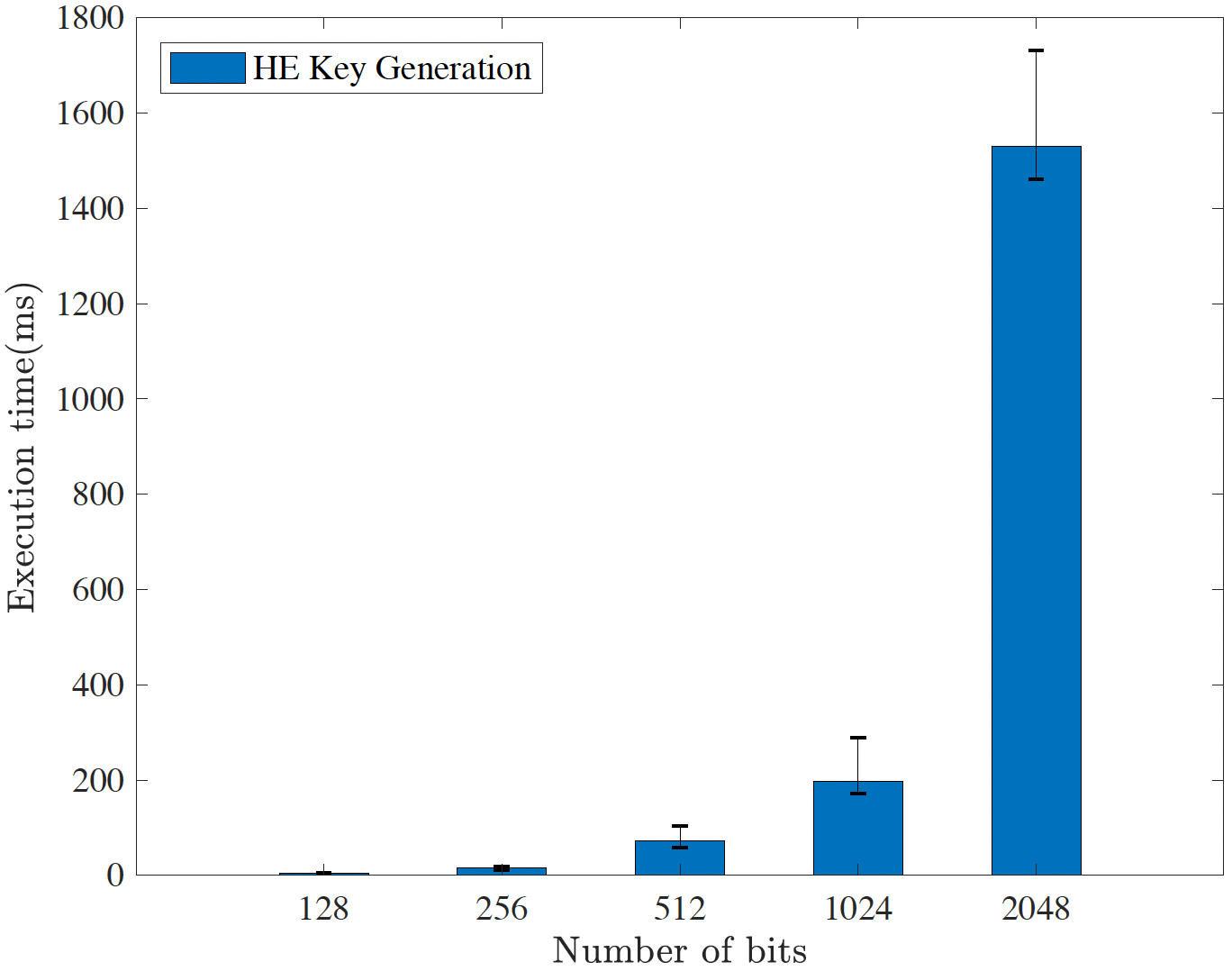}
\caption{Execution time of HE key generation process.}
\label{fig:hekey}
\end{figure}

{To evaluate distributed homomorphic-encrypted federated averaging cryptographic system performance, we tested the experiments based on jspaillier. We boosted the number of bits from 128 to 256, 512, 1024, and 2048 to measure the execution time of key pair generation procedures. As shown in Fig.~\ref{fig:hekey},  key generation time is 5 ms for 128 bits, 16 ms for 256 bits, 73 ms for 512 bits, 198 ms for 1024 bits, and 1531 bits for 2048 ms. It shows that key generation time will increase exponentially when the number of bits grows.
As we can see from the result, a 1,024-bit key pair needs the 512-bit prime number for the key generation process, which provides sufficient security requirements.}

\begin{figure}[t]
\centering
\includegraphics[width=0.24\textwidth]{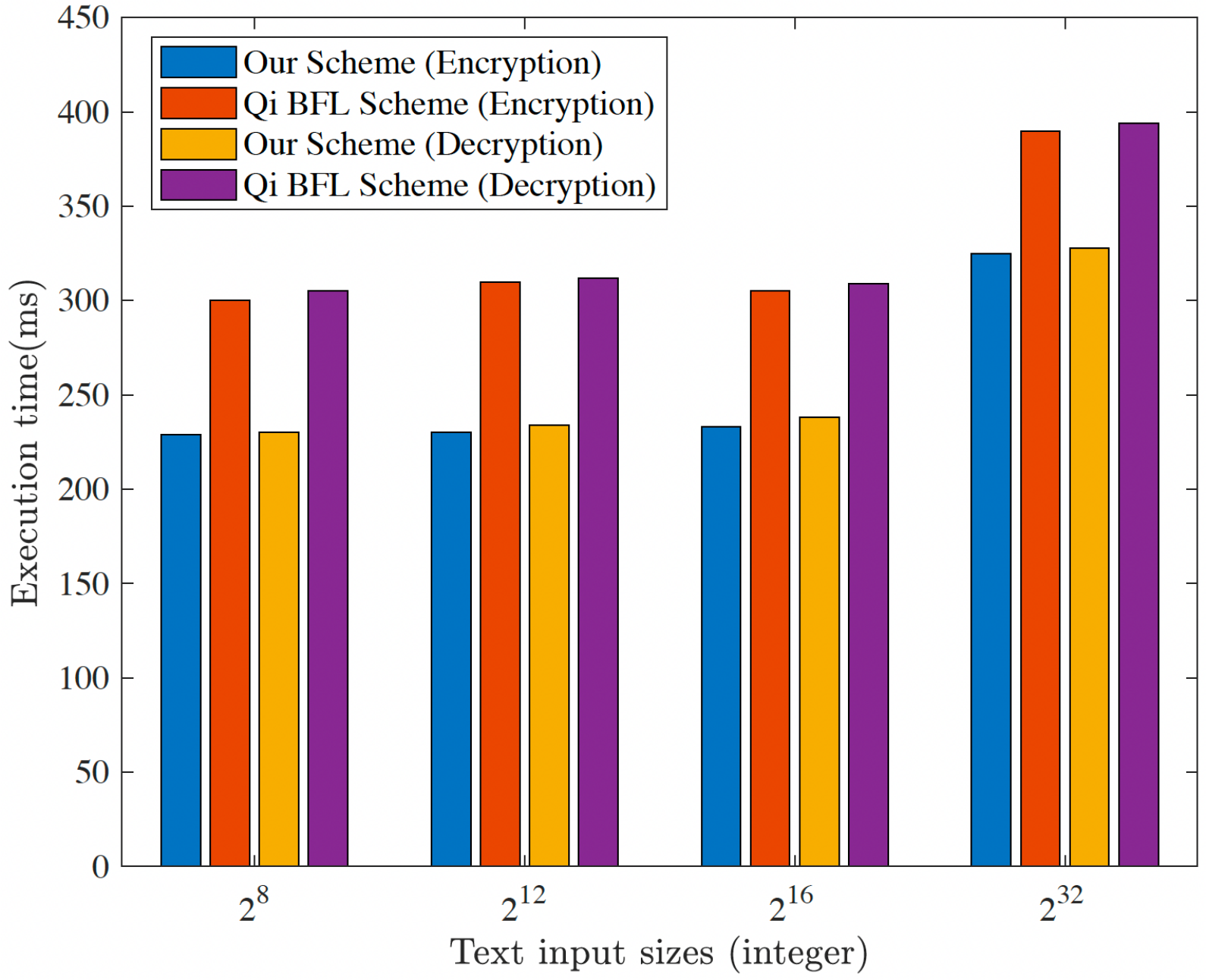}
\caption{Execution time comparison of our scheme, and Qi's scheme~\cite{qi2021blockchain}.}
\label{fig:heencdec}
\end{figure}




{Next, we compare the execution time of partial homomorphic addition operation ($(A+B)$), encrypted multiplication ($(A+B)*C$), and decryption ($(A+B)*C$) algorithms with Qi's scheme~\cite{qi2021blockchain}. As we can see from Fig.~\ref{fig:heencdec}, 
encrypted addition and decryption time for our scheme $2^{12}$ is 226 ms and 230 ms, which outperforms the  Qi's scheme~\cite{qi2021blockchain}. The encrypted addition and decryption time for $2^{32}$ are 307 ms and 325 ms in our scheme, and  Qi's scheme~\cite{qi2021blockchain} is around 400ms. Compared to other cases, $2^{32}$ case's execution cost all increases slightly. As a result, our proposed scheme could achieve better performance when increasing input data size for the traffic prediction task.}

 
 
 
\subsection{Federated Learning}
\label{sec:FL Exp}

\subsubsection{Experimental Setup}
The FL experiments are simulation-based and were conducted on Google Colab using one NVIDIA P100 GPU, two Intel(R) Xeon(R) CPUs @ 2.30GHz, and 13.34 gigabytes of RAM. {All experiments trained a multi-layer GRU (gated recurrent unit) model for predicting traffic flow at a 5-minute horizon. In our GRU model design, we stack two GRU layers in sequence with varying numbers of hidden units, followed by a dropout layer with the final layer as a densely connected output layer. To simulate online training, the input samples were dynamically fed in sequence during each FL round. At the start of each experiment, the parameters of $G^0$ are randomly initialized without a pre-trained model. Moreover, in addition to the federated models, we train a baseline reference model for comparative analysis. The baseline model architecture is the same as the federated model, but it is trained without federation exclusively using the data of a given individual detector.}

\subsubsection{Dataset and Study Area}
We use a real-world traffic flow dataset in our experiments. The Delaware Department of Transportation (DelDOT) provides the dataset and includes traffic flow data collected from DelDOT-maintained roadways at a 5-minute resolution. We select various nonsequential detectors along the I-95 north arterial to act as the FL clients in each experiment. Each client is provided a location-specific dataset containing point-based flow measurements from the start of August 2019 until the end of September 2019. We separate 80\% of the data for real-time training and inference while saving the remaining 20\% for future offline inference experiments.


\begin{algorithm}[t]
\caption{Online inference of client $c$ in round $r_i, i > 1$}
\label{alg:realtime_inference}
\SetAlgoLined
\LinesNumbered
\SetKwInOut{Input}{Input}
\SetKwInOut{Output}{Output}
\Input{$G^{i-1}$, $B^{i-1}_c$, $d^{old, i}_c$}
\Output{None}
\everypar={\nl}
\For{each client ${c \in \mathcal{C}}$ in region $e_i$ of $E$ during round $r_i$ of $R$ in parallel}
    {
        $d^{in, i}_c$, $BASE^{i}_c$, $FED^{i}_c$ $\leftarrow$ [], [], [];
        \Comment{Empty arrays.} \\
        
        $j$ $\leftarrow$ 0; \\
        
        $d^{pred, j}_c$ $\leftarrow$ $d^{old, i}_c[:input\_shape]$; \\
        \Comment{Extract $input\_shape$ number of the latest data.}
        
        \While{$j < input\_shape$}
        {
            $BASE^{i}_c$.\textsc{Add($c$.PredictBy($B^{i-1}_c$, $d^{pred, j}_c$));}
            
            $FED^{i}_c$.\textsc{Add($c$.PredictBy($G^{i-1}$, $d^{pred, j}_c$));}
            
            $d^{in, i}_c$.\textsc{Add($c$.CollectOneData())};
            
            $j \leftarrow j + 1$;
            
            $d^{pred, j}_c$ $\leftarrow$ $d^{pred, j}_c$.\textsc{PopLeft()} $\cup$ $d^{in, i}_c$;
        }
        $TRUE^{i}_c$ $\leftarrow$ $d^{in, i}_c$;
    }
\end{algorithm}

\subsubsection{Comparison with Other Blockchain-based Federated Learning Systems}

{In this subsection, we compare the different blockchain types, architecture, encryption methods, privacy protection, decentralized FedAvg, and the blockchain platform among our proposed scheme and other state-of-the-art blockchain-based federated learning systems, including the Li BFL \cite{bladefl2022}, Zhao BFL \cite{zhao2020privacy}, Fabric FL \cite{fabricfl2021}, BAFL \cite{fengbafl2022}, Biscotti BFL \cite{shayan2020biscotti}, Hierarchical BFL \cite{chai2020hierarchical}, and Qi BFL \cite{qi2021blockchain}.}

{As we observe from Table \ref{tab: consensus comparison}, most of the proposed blockchain-based federated learning systems utilize the permissioned blockchain type, and the one-layer blockchain architecture is popular. Only Qi BFL~\cite{qi2021blockchain},  and our proposed system offer partial HE features, and our scheme is based on the DHFA, which supports the decentralized FedAvg method. The bi-level blockchain architecture with DHFA encryption can protect data security in a more efficient and end-to-end privacy-preserving way. We provide detailed comparative experiments with Qi BFL~\cite{qi2021blockchain} in the HE performance evaluation section. For the Li BFL \cite{bladefl2022}, Zhao BFL \cite{zhao2020privacy}, Fabric FL \cite{fabricfl2021}, BAFL \cite{fengbafl2022}, Biscotti BFL \cite{shayan2020biscotti}, and Hierarchical BFL \cite{chai2020hierarchical}, we conduct the table comparison results since their evaluation is based on other cryptography techniques, FL simulation results, and other evaluation metrics. Most existing schemes utilize differential privacy requiring adding noise for the data. Our proposed DHFA scheme can solve the centralized FL server issue and perform calculations on the encrypted data.}

\begin{table*}[h]
\centering
\caption{Comparisons with other blockchain-based federated learning systems.}
\label{tab: consensus comparison}
\begin{tabular}{ccccccc}
Proposed Scheme & \thead{Blockchain Type} & \thead{Architecture} & \thead{Encryption method} & \thead{Privacy Protection} & \thead{Decentralized FedAvg} & Blockchain Platform \\ \hline
Li BFL \cite{bladefl2022}  & Permissionless       & One-layer  & Adding Noise & Differential Privacy & $\times$ & $\times$ \\ \hline
Zhao BFL \cite{zhao2020privacy}   & $\times$       & One-layer  & Adding Noise & Differential Privacy & $\times$ & $\times$ \\ \hline
Fabric FL \cite{fabricfl2021}  & Permissioned & One-layer  & $\times$ & $\times$ & Yes & Hyperledger Fabric \\ \hline
BAFL \cite{fengbafl2022}  & $\times$ &  One-layer & $\times$ & Entropy Weight &  $\times$ &  $\times$ \\ \hline
Biscotti BFL \cite{shayan2020biscotti}  & Permissioned &  One-layer & Secure Aggregation & Differential Privacy &  Yes &  $\times$ \\ \hline
Hierarchical BFL \cite{chai2020hierarchical}  & $\times$ & Hierarchical  & $\times$ & $\times$ & $\times$ & $\times$ \\ \hline
Qi BFL \cite{qi2021blockchain}  & Permissioned & One-layer  & Aggregation & Partial HE & $\times$ & Ethereum \\ \hline
Our Scheme  & Permissioned & Bi-level  & Secure FedAvg  & DHFA & Yes & Hyperledger Fabric \\ \hline
\end{tabular}
\end{table*}

\subsubsection{Regional Group Size Performance Comparison Analysis}
\label{sec:regional}
Motivated by our previous work in \cite{meese2022bfrt}, we select the GRU model as the chosen DL architecture for the FL experiments. Each FL simulation consists of 1165 rounds, where the duration of each round is one hour. In our dataset, sensors collect flow data at a 5-minute resolution, corresponding to twelve new data instances per hour per sensor. The GRU models are designed to process input instances sequentially, and one hour of historical flow data is used to generate the flow prediction for the next five minutes. As shown in Algo. \ref{alg:realtime_inference}, the sample size of $d^{in, i}_c, i > 1$ equals the $input\_shape$ of the GRU models (i.e., 12), to control data flow according to the hourly rounds. The exception is that $d^i_c.size, i = 1$ is set to 24 (i.e., 2 times of $d^{in, i}_c.size$, where $i > 1$) in $r_1$ for all ${c \in \mathcal{C}}$, because the models need at least 13 data samples for training. We set $MaxDataSize = 24$ in the regional group size performance analysis, providing the models with two hours of historical traffic flow data for training during each FL round. During simulation, $MaxDataSize$ controls the sample size of historical data to use when training for the current round. 
Moreover, we set the number of training epochs for each round to five.

\begin{figure*}[t]
\centering
\includegraphics[width=0.85\textwidth]{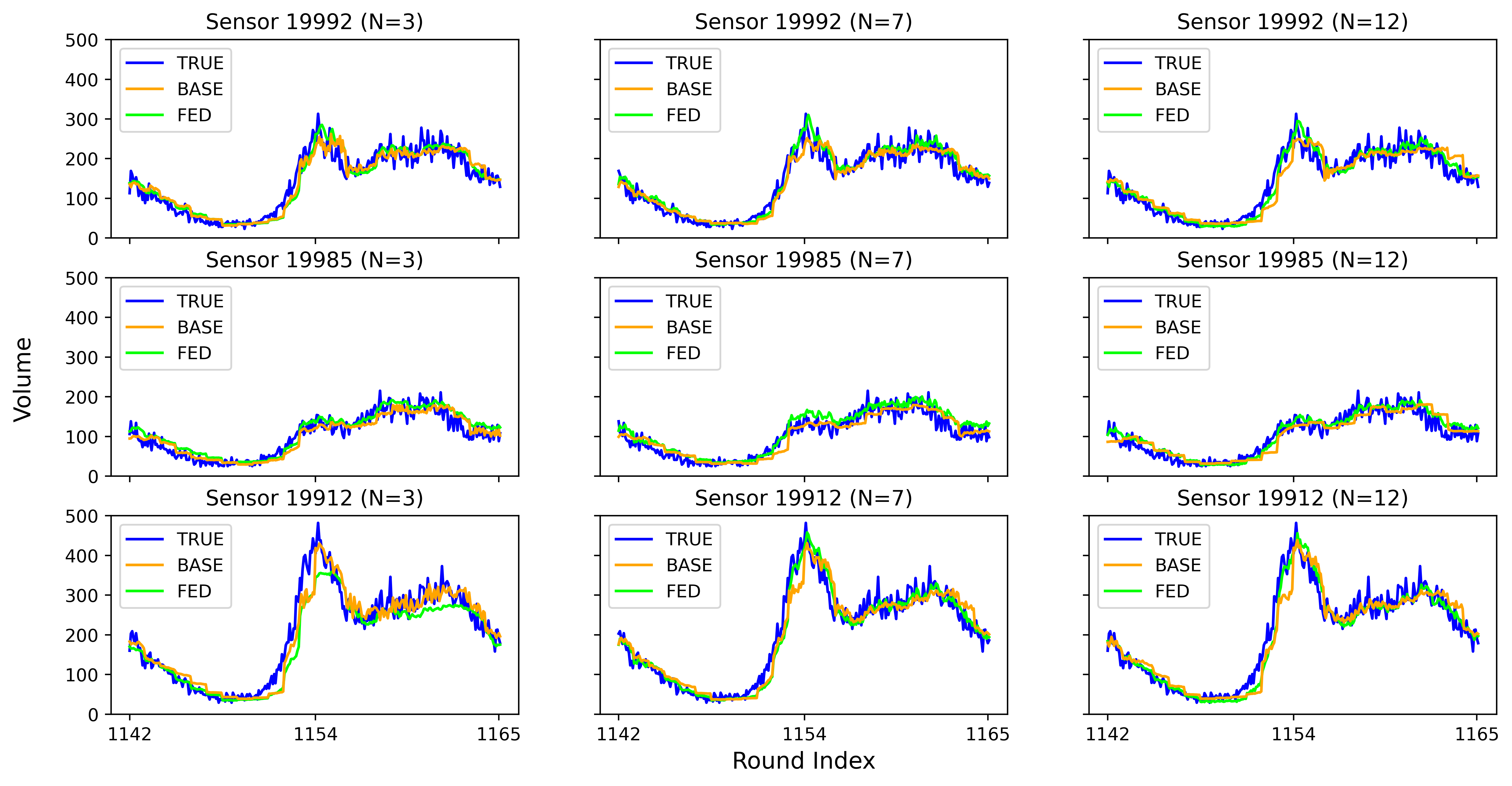}
\caption{Online prediction curves for the last 24 FL rounds with varying group sizes. The FL models are compared to the baseline cases for each reference detector.}
\label{fig:regionalfig}
\end{figure*}

\emph{B$^2$SFL} proposes a regional grouping of detectors within transportation networks to perform localized FL, producing region-specific online traffic prediction models. To elucidate the impacts of varying regional group sizes, we conduct FL simulations using identical GRU models while varying the number of participating clients (and, consequently, the number of distinct time series being considered). Fig. \ref{fig:regionalfig} illustrates the partial online inference curves for three traffic sensors. {We examine FL group sizes of 3, 7, and 12 detectors within a single region, together with comparison with the baseline case $B_c$ of a single detector. In the baseline case, the model is trained identically to the FL models, but DHFA is not invoked and the model is trained locally on a given REN and never leaves the device.} The plots compare the regional global model prediction output with the ground truth sensor data and the baseline case for the last 24 FL rounds, translating to a day of traffic flow data. The process for obtaining real-time inference values is shown in Algo. \ref{alg:realtime_inference}.
Specifically, for any client ${c \in \mathcal{C}}$ in round $r_i \in R, i > 1$, we generate a temporary dataset $d^{pred, j}_c$ by extracting $input\_shape$ of the most recent data points from $d^{old, i}_c$ (Algo. \ref{alg:realtime_inference}: line 4). 
Notably, $input\_shape = p$, where $p$ is the collection period for each FL round. We define $p = 12$ in this study, thus each $c$ collects one hour of new traffic data before training again. During data collection, each newly collected instance is appended to $d^{pred, j}_c$, while popping the oldest data instance to maintain a vector length of $input\_shape$ elements. 
After appending a new value, $c$ uses both $B^{i-1}_c$ and $G^{i-1}$ to perform online inference (Algo. \ref{alg:realtime_inference}: lines 7-11), updating $BASE^{i}_c$ and $FED^{i}_c$ with the resulting prediction. After completing an $input\_shape$ number of prediction steps, the resulting $BASE^{i}_c$ and $FED^{i}_c$ will contain the inference values from the corresponding models for $r_i, i > 1$. Lastly, $d^{in, i}_c$ is assigned to $TRUE^{i}_c$ (Algo. \ref{alg:realtime_inference}: line 13), which represents the \emph{TRUE} curve(s) in Fig. \ref{fig:regionalfig}.

{Examining Fig. \ref{fig:regionalfig} indicates that the group size directly impacts the resulting global model prediction accuracy. For example, comparing the $N=3$ and $N=7$ groups, we can see that increasing the group size to seven improved the prediction performance for both sensors 19992 and 19912, with 19912 experiencing the most improvement. On the other hand, the global model prediction accuracy for sensor 19985 suffered when the group size was increased. Notably, despite the spatial closeness of the three sensors in the $N=3$ group, as shown in Fig. \ref{fig:regionalfig}, we can see that the time series distribution differs in trend and magnitude between the sensors. Consequently, the prediction accuracy degrades when these sensors are grouped in a small regional cluster due to the federated averaging process for merging the parameters. }


{To view the differences in accuracy at a more granular level, Table \ref{table:regionalcompare} provides the mean absolute error (MAE), mean squared error (MSE), root mean squared error (RMSE), and mean absolute percent error (MAPE) for three representative detectors present in the three experimental groups ($N={3, 7, 12}$), while the case $N=1$ represents $B_c$, the baseline model without federation.} Notably, we can see that a group size of seven resulted in the best performance for detectors 19912 and 19992, and the performance was significantly improved from the baseline and $N=3$ cases. However, when further increasing the group size to 12, we see a decrease in performance across these two detectors. These results reveal that controlling for group size has an important impact on accuracy.


\begin{table}[]
\caption{Prediction error calculations under various metrics and group sizes ($N$)}
\label{table:regionalcompare}
\centering
\begin{tabular}{|c|c|c|c|c|}
\hline
\textbf{Metric}                 & \textbf{Group Size}           & \textbf{19912}                 & \textbf{19985}                 & \textbf{19992}                 \\ \hline
                                & N = 1                         & 24.76                          & 15.33                          & 18.98                          \\ \cline{2-5} 
                                & N = 3                         & 28.12                          & 16.81                          & 18.97                          \\ \cline{2-5} 
                                & \cellcolor[HTML]{9AFF99}N = 7 & \cellcolor[HTML]{9AFF99}19.79  & 17.20                          & \cellcolor[HTML]{9AFF99}16.72  \\ \cline{2-5} 
\multirow{-4}{*}{\textbf{MAE}}  & N = 12                        & 20.29                          & \cellcolor[HTML]{9AFF99}15.16  & 17.72                          \\ \hline
                                & N = 1                         & 1171.76                        & 396.44                         & 654.09                         \\ \cline{2-5} 
                                & N = 3                         & 1666.78                        & 434.75                         & 621.95                         \\ \cline{2-5} 
                                & \cellcolor[HTML]{9AFF99}N = 7 & \cellcolor[HTML]{9AFF99}717.44 & 479.97                         & \cellcolor[HTML]{9AFF99}489.48 \\ \cline{2-5} 
\multirow{-4}{*}{\textbf{MSE}}  & N = 12                        & 753.69                         & \cellcolor[HTML]{9AFF99}360.59 & 512.59                         \\ \hline
                                & N = 1                         & 34.22                          & 19.89                          & 25.47                          \\ \cline{2-5} 
                                & N = 3                         & 40.83                          & 20.85                          & 24.94                          \\ \cline{2-5} 
                                & \cellcolor[HTML]{9AFF99}N = 7 & \cellcolor[HTML]{9AFF99}26.79  & 21.91                          & \cellcolor[HTML]{9AFF99}22.12  \\ \cline{2-5} 
\multirow{-4}{*}{\textbf{RMSE}} & N = 12                        & 27.45                          & \cellcolor[HTML]{9AFF99}18.99  & 22.64                          \\ \hline
                                & N = 1                         & 0.15                           & \cellcolor[HTML]{FFFC9E}0.18   & 0.16                           \\ \cline{2-5} 
                                & N = 3                         & 0.15                           & 0.21                           & 0.16                           \\ \cline{2-5} 
                                & \cellcolor[HTML]{9AFF99}N = 7 & \cellcolor[HTML]{9AFF99}0.12   & 0.20                           & \cellcolor[HTML]{9AFF99}0.14   \\ \cline{2-5} 
\multirow{-4}{*}{\textbf{MAPE}} & N = 12                        & 0.13                           & \cellcolor[HTML]{FFFC9E}0.18   & 0.15                           \\ \hline
\end{tabular}
\end{table}

\begin{table}[]
\caption{Online inference errors during the last 24 FL rounds for the light-weight model}
\centering
\label{table:lightweight}
\begin{tabular}{|c|c|c|c|c|}
\hline
\textbf{Detector} & \textbf{MAE} & \textbf{MSE} & \textbf{RMSE} & \textbf{MAPE} \\ \hline
19912             & 31.28        & 1745.70      & 41.78         & 0.19          \\ \hline
19924             & 81.13        & 11199.60     & 105.83        & 0.24          \\ \hline
19951             & 51.51        & 5036.58      & 70.97         & 0.19          \\ \hline
19978             & 27.24        & 1053.15      & 32.45         & 0.56          \\ \hline
19985             & 18.02        & 548.59       & 23.43         & 0.23          \\ \hline
19992             & 21.77        & 805.77       & 28.39         & 0.20          \\ \hline
19997             & 34.30        & 2183.80      & 46.73         & 0.18          \\ \hline
\end{tabular}
\end{table}

\begin{figure*}[t]
\centering
\includegraphics[width=0.82\textwidth]{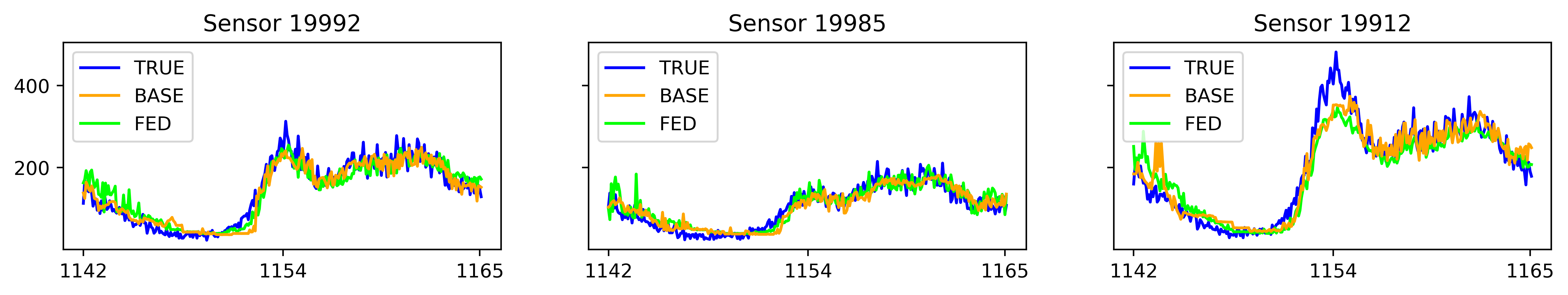}
\caption{Online prediction curves for the last 24 FL rounds using the reduced parameter model. The FL models are compared to the baseline cases for each reference detector.}
\label{fig:lightweight_plot}
\end{figure*}

\begin{figure}[t]
\centering
\includegraphics[width=0.36\textwidth]{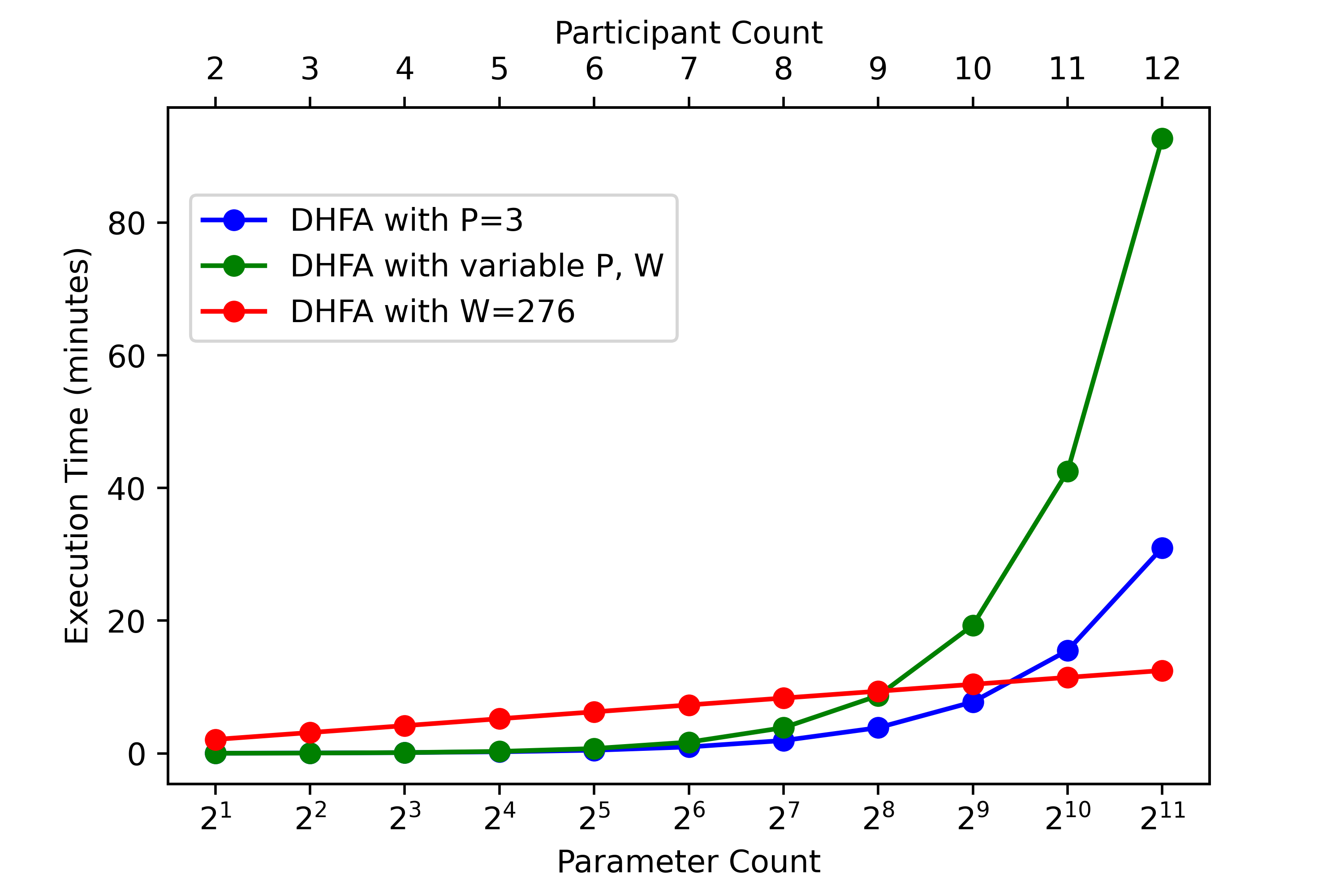}
\caption{Execution time complexity model for the proposed DHFA algorithm.}
\label{fig:execution_time_dhfa}
\end{figure}

\subsubsection{Reduced Parameter Model Performance Comparison Analysis}

In this experiment, we analyze the online traffic prediction accuracy of a lightweight GRU model having considerably lower parameters than existing approaches \cite{liu2020privacy}. While DHFA can be expensive to execute for deep models, recent research has highlighted that many of the learned parameters provide little impact on the prediction output \cite{cheng2017survey}. Motivated by this observation, this experiment assesses the effects of drastically reducing the number of parameters in our FL-based online traffic flow prediction (TFP) model to reduce DHFA execution time. 


{When building the model, we instantiate seven participants using the same detectors and data set as the $N = 7$ group from Sec. \ref{sec:regional}. The GRU model structure is the same, with two sequential GRU layers. However, each layer's hidden units is reduced to 5 (from 50 previously). We also set $MaxDataSize = 240$ and the number of epochs to fifty during training. Under this design, the number of trainable model parameters is reduced from 23,001 to 276. The FL simulation is conducted in the same manner mentioned in Sec. \ref{sec:regional}. Fig. \ref{fig:lightweight_plot} illustrates the real-time comparative prediction curve for detectors 19992, 19985, and 19912, while Table \ref{table:lightweight} provides the prediction errors for each detector during the last 24 simulated communication rounds.}

Using detector 19992 as a comparative reference, we can see that the MAPE prediction error increased from 0.14 to 0.20 when reducing the number of learnable parameters, representing a 42.86\% change in accuracy. While this increase is substantial, it is notable that the reduction in learnable parameters between the two models represents a 98.8\% decrease in weights.




\subsubsection{Execution Time Complexity Model}
This subsection presents the execution time complexity model for the proposed DHFA within our online traffic flow prediction workflow using GRU. Notably, DHFA performs two fundamental operations: encrypted addition and encrypted multiplication. During execution, the parameters of each participant in a given regional FL group will be securely averaged together using both operations. Consequently, the number of operations performed during each execution is a function of two parameters: $W$, representing the model's total number of learnable weights, and $P$, denoting the number of participants in a given regional FL group. Computing $W$ is inherently specific to the model architecture, and in this analysis, we focus on the GRU model as an example. The GRU model is a recurrent neural network (RNN) consisting of three feedforward neural networks (FFNN) structured as a series of gates. The following equation can be used to calculate the number of trainable weights within a single FFNN:
\begin{align*}
    W_i = (h(h+i) + h)
\end{align*}

where $h$ indicates the size of the hidden layer and $i$ denotes the length of the input vector. Similarly, because GRU consists of 3 FFNNs, we can compute the total number of learnable parameters within a single GRU layer with the equation:

\begin{align*}
    W = \sum_{n=1}^{3} W_i
\end{align*}


{For the first layer, $i_1=1$, because the traffic sequence is fed into the model sequentially. In the subsequent GRU layer, $i_2=5$ represents the number of hidden units in the first layer, which is five in our lightweight model. Our output layer in all models is a fully connected layer, where $W_i= i_2 + 1$ accounts for the hidden representation for $i_2$ neurons and the final output weight. Consequently, the total parameter count can be computed for our lightweight model using the presented equations to determine that $W = 276$.}

For modeling the execution time of DHFA, we use the experimental values presented in Sec. \ref{section:HE cost} for an input data size of $2^{12}$: 226ms for encrypted addition and 230ms for encrypted multiplication. During execution, DHFA will compute $P-1$ additions and one multiplication for each of the $W$ parameters in the model. Accordingly, the total execution time $T$ can be estimated using the following equation:

\begin{align*}
    T = W(226*(P-1)+230)
\end{align*}

The results for various values of $W$ and $P$ are illustrated in Fig. \ref{fig:execution_time_dhfa}. Within the figure, three calculations are presented: (1) the blue line computes $T$ with a static $P=3$; (2) the green line computers $T$ with variable $P$ and $W$ at each point; and (3) the red line computes $T$ with a static $W=276$ corresponding to the value of $W$ for our lightweight model. Notably, the bottom x-axis is scaled exponentially. {In comparison, the unencrypted federated averaging algorithm generally runs on the order of milliseconds due to its simplicity. Consequently, integrating DHFA is computationally expensive compared to a system without encryption and is suitable for privacy-centric ITS applications.}



\section{Conclusion}

{This paper proposed a bi-level blockchain architecture for secure federated learning-based traffic prediction.
The bottom and top layer blockchains store local and aggregated global parameters. We design the partial private key distribution protocol and the partially homomorphic encryption scheme to achieve the privacy-preserving federated averaging procedure. We conducted both system correctness and security discussions to validate our design. We implemented the proposed architecture by utilizing Hyperledger Fabric, Jspaillier library, and the Google Colab platform. The experiment results indicate that the proposed scheme is secure and efficient for decentralized federated averaging schemes and real-world traffic prediction tasks.}

\ifCLASSOPTIONcaptionsoff
  \newpage
\fi

\bibliographystyle{IEEEtran}
\bibliography{bibtex/IEEEabrv,bibtex/sig}


\vskip -2\baselineskip plus -1fil 
\begin{IEEEbiography}
[{\includegraphics[width=1.0in,height=1.25in,clip]{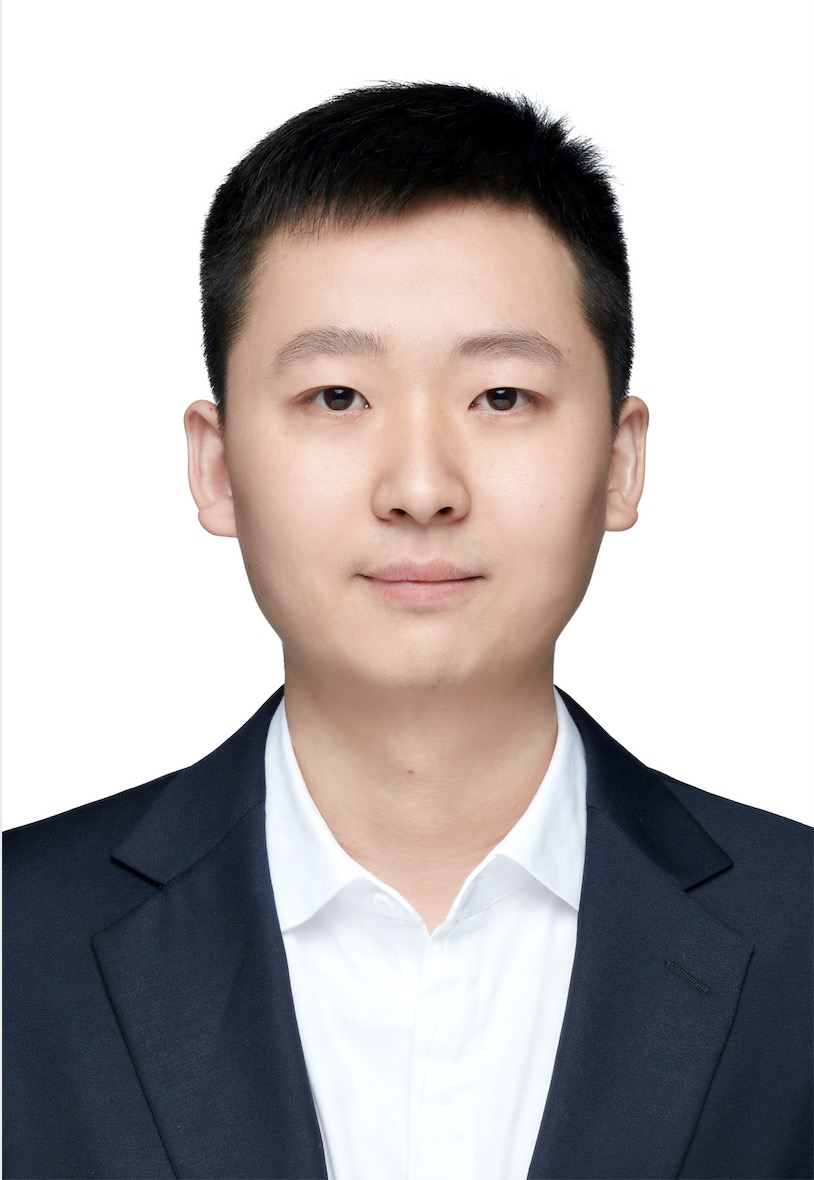}}]{Hao Guo} received the B.S. and M.S. degrees from the Northwest University, Xi'an, China in 2012, and the Illinois Institute of Technology, Chicago, United States in 2014, and his Ph.D. degree from the University of Delaware, Newark, United States in 2020, all in computer science.
He is currently an Assistant Professor
with the School of Software at
the Northwestern Polytechnical University.
His research interests include blockchain and distributed ledger technology, data privacy and security, cybersecurity, cryptography technology, and Internet of Things (IoT). He is a member of both ACM and IEEE.
\end{IEEEbiography}

\vskip -6pt plus -1fil

\begin{IEEEbiography}
[{\includegraphics[width=1in,height=1.25in,clip]{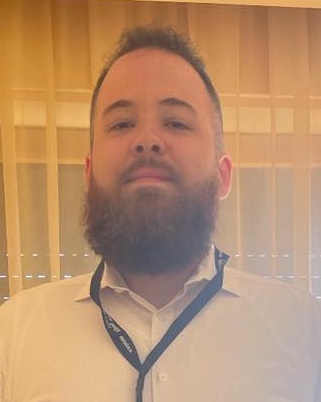}}]{Collin Meese} received his B.S. degree in computer science from the University of Delaware in 2020. He is currently working toward the Ph.D. degree at the University of Delaware. His research interests include blockchain, vehicular networks, distributed and high-performance computing, connected and autonomous vehicles, and intelligent civil systems. He is a student member of IEEE. 
\end{IEEEbiography}

\vskip -6pt plus -1fil
\begin{IEEEbiography}
[{\includegraphics[width=1in,height=1.25in,clip]{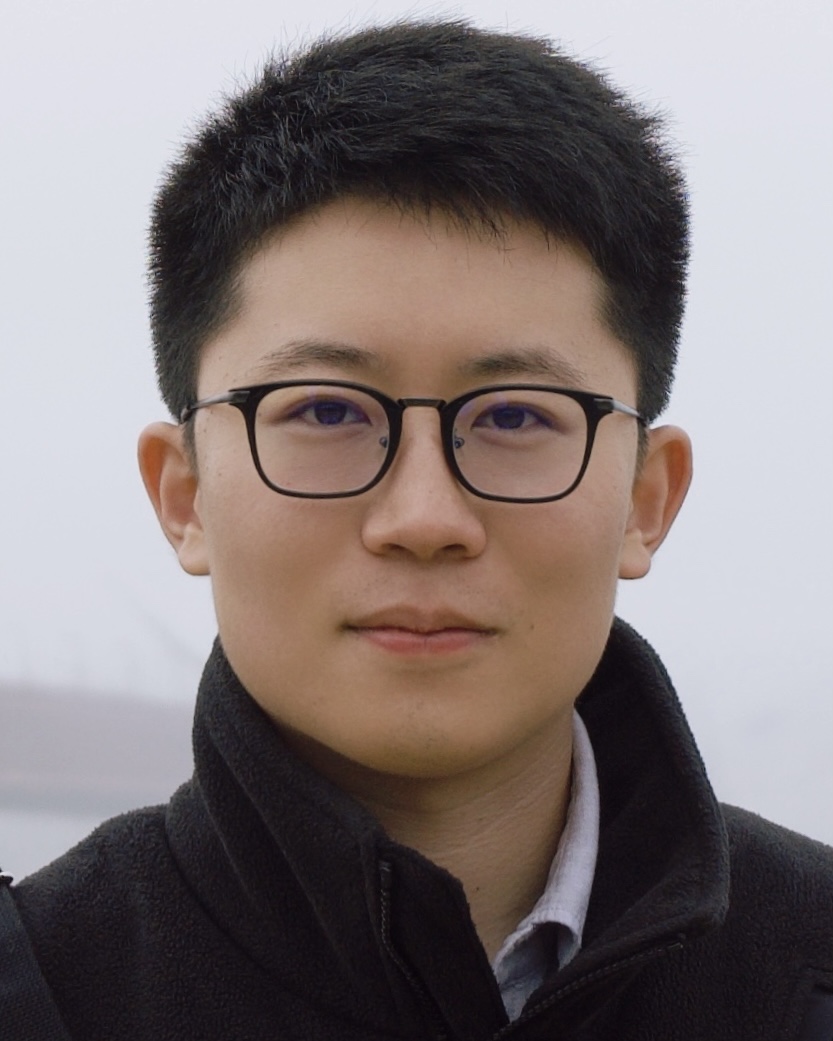}}]{Wanxin Li} is a Lecturer in the Department of Communications and Networking at Xi'an Jiaotong-Liverpool University. He received his B.S. degree from Chongqing University in 2015, and his M.S. and Ph.D. degrees from the University of Delaware in 2017 and 2022, respectively. He was a recipient of IEEE ITSS Best Dissertation Award and IEEE TEMS Outstanding Ph.D. Dissertation Award in 2022. His research interests include blockchain, cryptography, distributed AI, and smart transportation. He is a member of IEEE and ACM.
\end{IEEEbiography}
\vskip -6pt plus -1fil
\begin{IEEEbiography}[{\includegraphics[width=1in,height=1.125in,clip]{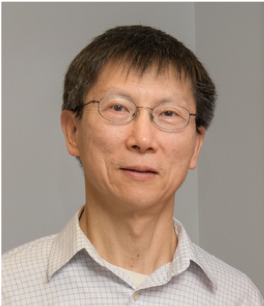}}]{Chien-Chung Shen} received his B.S. and M.S. degrees from National Chiao Tung University, Taiwan, and his Ph.D. degree from UCLA, all in computer science.  He was a research scientist at Bellcore Applied Research working on the control
and management of broadband networks.  He is now a Professor in the Department of Computer and Information Sciences at the University of Delaware. His research interests include blockchain, federated learning, Wi-Fi, SDN, digital twins, and cybersecurity education. He is a recipient of the NSF CAREER Award and a member of ACM and IEEE.
\end{IEEEbiography}
\vskip -6pt plus -1fil

\begin{IEEEbiography}
[{\includegraphics[width=1in,height=1.25in,clip]{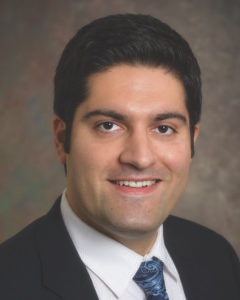}}]{Mark Nejad} is an Assistant Professor at the University of Delaware. His research interests include  network optimization, distributed systems, blockchain, game theory, and automated vehicles. He has published more than forty peer-reviewed papers and received several publication awards including the 2016 best doctoral dissertation award of the Institute of Industrial and Systems Engineers (IISE) and the 2019 CAVS best paper award IEEE VTS. His research is funded by the National Science Foundation and the Department of Transportation. He is a member of the IEEE and INFORMS.
\end{IEEEbiography}







\end{document}